\documentclass[12pt,epsf,amssymb,ulem,qsymbols]{article}
\usepackage{tabularx}
\usepackage{array}
\usepackage{graphics}
\usepackage{graphicx}
\usepackage{psfrag}
\usepackage{epsfig}
\usepackage{amsmath}
\usepackage{amssymb}
\usepackage{ulem}
\usepackage{setspace}
\usepackage{rotating}
\usepackage{colortbl}
\usepackage{tabularx}
\usepackage{longtable}
\usepackage{lineno}
\makeatletter
\usepackage{textcomp}
\usepackage[usenames,dvipsnames]{xcolor}
\usepackage{relsize}

\usepackage{verbatim}

\setlongtables

\newcommand{\msbar}{{\overline{\rm MS}}}

\newcommand{\bea}{\begin{eqnarray}}
\newcommand{\eea}{\end{eqnarray}}
\newcommand{\beq}{\begin{equation}}
\newcommand{\eeq}{\end{equation}}
\newcommand{\ec}{\end{center}}
\newcommand{\bc}{\begin{center}}
\newcommand{\Bdecpi}{{\rm B}^{0} \rightarrow \pi^- e^+ \nu_e }
\newcommand{\gev}{{\rm GeV}}
\newcommand{\mev}{{\rm MeV}}

\newcommand{\pdir}{p\kern -5.2pt\raise 0.2ex\hbox {/}}

\newcommand{\vdir}{v\kern -5.75pt\raise 0.15ex\hbox {/}}
\newcommand{\kdir}{k\kern -5.75pt\raise 0.15ex\hbox {/}}
\newcommand{\epsdir}{\epsilon\kern -5.0pt\raise 0.15ex\hbox {/}}
\newcommand{\bvdir}{\bar{v}\kern -5.75pt\raise 0.15ex\hbox {/}}
\newcommand{\Ddir}{D\kern -7.75pt\raise 0.20ex\hbox {/}}
\newcommand{\Adir}{A\kern -7.75pt\raise 0.20ex\hbox {/}}
\newcommand{\ldir}{l\kern -5.0pt\raise 0.2ex\hbox{/}}
\newcommand{\varepsdir}{\varepsilon\kern -5.5pt\raise 0.15ex\hbox{/}}

\newcommand{\nf}{{N_{\rm f}}}

\newcommand{\nn}{\nonumber}
\makeatother

\begin{document}
\thispagestyle{empty} 
\begin{flushright}
\begin{tabular}{l}
{\tt \footnotesize LPT 14-58}\\
\end{tabular}
\end{flushright}
\begin{center}
\vskip 2.8cm\par
{\par\centering \textbf{\LARGE  
\Large \bf Insight into  $D/B\to \pi \ell \nu_\ell$ decay using}}\\
\vskip .35cm\par
{\par\centering \textbf{\LARGE  
\Large \bf models with an effective pole}}\\
\vskip 1.05cm\par
{\scalebox{.8}{\par\centering \large  
\sc D. Be\v{c}irevi\'c$^a$, A. Le Yaouanc$^a$, A. Oyanguren$^b$, P. Roudeau$^c$ and F. Sanfilippo$^{d}$}
{\par\centering \vskip 0.65 cm\par}
{\sl 
$^a$~Laboratoire de Physique Th\'eorique (B\^at.~210)\\
Universit\'e Paris Sud and CNRS (UMR 8627), F-91405 Orsay-Cedex, France.}\\
{\par\centering \vskip 0.25 cm\par}
{\sl 
$^b$~IFIC, Universidad de Val\`encia - CSIC, \\
Cat\'edratico Jos\'e Beltr\'an, 2, E-46980 Paterna, Spain.}\\
{\par\centering \vskip 0.25 cm\par}
{\sl 
$^c$~Laboratoire de l'Acc\'el\'erateur Lin\'eaire, \\
Universit\'e Paris Sud and CNRS (UMR 8607), F-91405 Orsay-Cedex, France.}\\
{\par\centering \vskip 0.25 cm\par}
{\sl 
$^d$~School of Physics and Astronomy, University of Southampton\\
Southampton SO17 1BJ, UK.}\\ 

{\vskip 1.65cm\par}}
\end{center}

\vskip 0.55cm
\begin{abstract}
After improving the knowledge about residua of the semileptonic form factor at its first two poles we show that they do not saturate the experimental data for $f_+^{D\pi}(q^2)$. To fill the difference we approximate the rest of discontinuity by an effective pole and show that the data can be described very well with the position of the effective pole larger than the next excitation in the spectrum of $D^\ast$ state. The results of fits with experimental data also suggest the validity of superconvergence which in the pole models translates into a vanishing of the sum of residua of the form factor at all poles. A similar discussion, complemented with the heavy quark scaling, can be employed in the case of $B\to \pi \ell\nu_\ell$. This also opens a possibility of extracting $\vert V_{ub}\vert$, the error of which appears to be dominated by $g_{B^\ast B\pi}$, that can be nowadays computed on the lattice. In evaluating the residua of the form factors at their nearest pole we needed the vector meson decay constants $f_{D^\ast}$ and $f_{B^\ast}$, which we computed by using the numerical simulations of QCD on the lattice with $N_{\rm f}=2$ dynamical quarks. We obtain, $f_{D^\ast}/f_D=1.208(27)$ and $f_{B^\ast}/f_B=1.051(17)$.
\end{abstract}
\newpage
\setcounter{page}{1}
\setcounter{footnote}{0}
\setcounter{equation}{0}
\noindent

\renewcommand{\thefootnote}{\arabic{footnote}}

\setcounter{footnote}{0}
\section{\label{sec-0}Introduction}
Over the past several years we witnessed a substantial progress in measuring  the differential decay rate of the semileptonic decay $D\to \pi \ell\nu_\ell$, with $\ell \in \{e,\mu\}$, which then allows us to study the shape of the relevant hadronic form factor since
\bea\label{eq:ddr}
{d\Gamma(D\to \pi\ell\nu_\ell)\over dq^2}= {G_F^2\vert V_{cd}\vert^2\over 192 \pi^3 m_D^3} \lambda^{3/2}(q^2,m_D^2,m_\pi^2) \vert f_+(q^2)\vert^2\,,
\eea
where the semileptonic form factor $f_+(q^2)$ is a function of $q^2=(p_D-p_\pi)^2$, and $\lambda(q^2,m_D^2,m_\pi^2)=[(q^2-(m_D-m_\pi)^2] [(q^2-(m_D+m_\pi)^2]$.  Throughout this paper we will use $\vert V_{cd}\vert =0.2252$, fixed from the unitarity of the Cabibbo--Kobayashi-Maskawa matrix~\cite{PDG}. On the basis of analyticity of the matrix element we assume that the form factor satisfies the unsubtracted dispersion relation, 
\bea\label{eq:unstr}
f_+(q^2) = {1\over \pi}\int \displaylimits_{t_0}^\infty { {\rm Im}f_+(t)\over t-q^2-i\varepsilon} dt\,,
\eea
which relates the semileptonic form factor defined for $q^2\in [m_\ell^2, (m_D-m_\pi)^2]$, to the discontinuity along the cut starting from $t_0=(m_D+m_\pi)^2$ where the $\ell \nu \to D\pi$ channel is opened. 
In the $B$-decay case, which we also consider, there is also an isolated pole below $t_0$.  The use of an unsubtracted dispersion relation for this form factor is actually allowed by perturbative QCD where 
for large space-like $Q^2=-q^2$ one finds that~\cite{Floratos:1978jb} 
\bea\label{pertQCD}
f_+(Q^2) \propto {\alpha_s(Q^2)\over Q^2}\,,
\eea
thus falling (slightly) faster than $1/Q^2$. Another important consequence of this behaviour is superconvergence that will be discussed in sec.~\ref{super}.

The formidable task of describing  the physical cut, ${\rm Im}f_+(t)$,  is often simplified by replacing it by a sum of poles, i.e. states with $J^P=1^-$ that couple to the vector current $\bar d \gamma_\mu c$ which mediates the semileptonic $D\to \pi$ decay. It can be replaced exactly by an infinite series of poles in quenched QCD or in the limit of $N_c\to \infty$, where all the states are infinitely narrow. This amounts to replacing the above integral by 
\bea\label{eq:s1}
f_+(q^2) = \sum_{n=0}^\infty { \; \displaystyle{\underset{q^2=m_{D_n^\ast}^2}{\rm Res}} f_+(q^2)\;\over m_{D^\ast_n}^2-q^2} \,,
\eea
with the residua given by 
\bea\label{eq:s2}
 \displaystyle{\underset{q^2=m_{D_n^\ast}^2}{\rm Res}} f_+(q^2)= {1\over 2} m_{D_n^\ast} f_{D_n^\ast} g_{D_n^\ast D\pi}\,, 
\eea
where we adopted the usual definitions of the vector meson decay constant $f_{D_n^\ast}$ and the pionic coupling $g_{D_n^\ast D\pi}$, namely
\bea
&&\langle D_n^\ast (p,\lambda)\vert \bar d\gamma_\mu c \vert 0 \rangle = m_{D_n^\ast} f_{D_n^\ast} \epsilon_\mu^\lambda \,,\nn\\
&& \cr
&& \langle D(p-k) \pi^\pm (k) \vert D_n^\ast (p,\lambda)\rangle = (k\cdot \epsilon)\ g_{D_n^\ast D\pi}\ ,
\eea
$\epsilon_\mu^\lambda$ being the polarization vector of the $D_n^\ast$ meson. 
In defining the coupling $g_{D_n^\ast D\pi}$ we assumed the pion to be charged. The case of the neutral pion is simply related to the charged one via isospin, e.g. $g_{D_n^\ast D\pi}\equiv g_{D_n^{\ast +} D^0 \pi^+}=  \sqrt{2} g_{D_n^{\ast +} D^+ \pi^0}$. 
In practice, this representation is corrected by loop effects, which we briefly discuss in sec.~\ref{loops}. Note that in eq.~(\ref{eq:s1}), $n$ labels the whole series of vector states starting from $m_0=m_{D^\ast}$, lowest lying vector meson (which is not to be confused with the main quantum number).  

Knowing all the couplings and decay constants is practically not feasible and one then tries truncating the infinite series. The simplest approximation, known as the vector meson dominance (VMD), consists in abruptly truncating the above eq.~(\ref{eq:s1}) by retaining only the first term,
\bea
f_+(q^2) = {1\over 2 m_{D^\ast}} {\quad f_{D^\ast} g_{D^\ast D\pi}\quad  \over 1- \displaystyle{q^2/ m_{D^\ast}^2} }\,.
\eea
That approximation turns out to be inadequate to describe the actual behavior of the $D\to \pi \ell\nu$ form factor. If the mass of the nearest pole and the value of the residuum were left as free parameters of the fit, the above relation could be used to fit the data. However, since we know the value of the position of the nearest pole,~\footnote{$m_{D^{\ast 0}}= 2.007~\gev$, $m_{D^{\ast \pm}}= 2.010~\gev$.}  it can be easily seen that the above VMD form does not reproduce the shape of the form factor. It is therefore necessary to include extra poles, or an extra {\sl effective} pole, the role of which is to approximate the cut beyond the first pole. Together with the arguments based on heavy quark symmetry relations, this was a starting point to build a parameterization proposed in ref.~\cite{BK}. Today, we have more information not only about the position of the first pole, 
but also about its residuum. Furthermore, the position of the second pole in the sum (\ref{eq:s1}) is also known, and we can get a reasonably good information about the residuum of the form factor $f_+^{D\pi}(q^2)$ at that (second) pole $t=m_{D^{\ast \prime}}^2$. 
In this paper we will improve on the discussion made in ref.~\cite{BK} and include the available information about the first two poles while approximating the rest of the discontinuity, ${\rm Im}f_+(t)$, by an effective pole. We will then compare the mass of that effective pole to the values obtained from a (reliable) quark model estimates, namely~\cite{GI}
\bea\label{gid}
n=1,\quad  L=0 &:&  m_{D^\ast} = 2.037~\gev\,,\cr
n=2,\quad  L=0 &:&  m_{D^{\ast\prime }} = 2.645~\gev\,,\cr
n=1,\quad  L=2 &:&  m_{D^{\ast\ast}} = 2.816~\gev\,,\cr
n=3,\quad  L=0 &:&  m_{D^{\ast ''}} = 3.11~\gev\,,
\eea
where $n$ is the main quantum number ordering the radial excitations, while $L$ orders the orbital excitations. Notice that the $m_{D^\ast}$ value obtained by the quark model~\cite{GI} is slightly larger than the experimentally established one. One can then envisage shifting the whole spectrum to reproduce the correct   $m_{D^\ast}^{\rm exp}$. We should emphasize that the state between the first and the second radial excitation is the orbitally excited one with $L=2$ but $n=1$. Indeed, a recent experimental $D$-meson spectroscopy analysis at LHCb~\cite{LHCb3} showed a presence of the state with 
$m_1=m_{D^{\ast\prime }} =2.649(5)$~GeV and the width $\Gamma_1=140(25)$~MeV, and another one with $m_2=2.761(8)$~GeV and a smaller width $\Gamma_2=74(37)$~MeV. A true second radial excitation is expected to lie beyond that latter state.  The result for $m_{D^{\ast\prime }}$ is obviously consistent with the above quark model prediction but  larger than the value reported by BaBar in ref.~\cite{delAmoSanchez:2010vq},   
$m_{D^{\ast\prime }} =2.609(4)$~GeV and $\Gamma_1=93(14)$~MeV. The BaBar value instead would be more consistent with the value in~(\ref{gid}) after shifting the spectrum to reproduce the correct  $m_{D^\ast}^{\rm exp}$, i.e. $30$~MeV lower than $m_1$ given in eq.~(\ref{gid}).

The remainder of this paper is organized as follows: In sec.~\ref{sec-res1} we discuss the value of the residuum of the form factors $f_+^{D\pi}(q^2)$ and $f_+^{B\pi}(q^2)$ at its nearest pole respectively. In that section we present the new lattice QCD determination of the ratios of decay constants $f_{D^\ast}/f_D$ and $f_{B^\ast}/f_B$. Constraints on the second pole residuum is discussed in sec.~\ref{sec-res2}. In sec.~\ref{super} we parameterize the remaining contributions to the form factor by an effective pole and evoke the superconvergence condition that can constrain the residuum of the effective pole. The ideas presented in the previous sections are then confronted with the actual experimental data in sec.~\ref{fitting} where the initial discussion of the $f_+^{D\pi}(q^2)$ form factor is extended to the $f_+^{B\pi}(q^2)$ case which leads to an estimate of $|V_{ub}|$. We briefly summarize in sec.~\ref{sec:concl}. In two appendices we give detailed results of the ratio of vector and pseudoscalar heavy-light meson decay constants computed on our lattices, and provide a discussion of the fit used to describe the experimental data paying particular attention to correctly accounting for the systematic uncertainties.

\section{Residuum of the form factor at its nearest pole\label{sec-res1}}

In this section we will discuss the value of the residuum of the form factor $f_+^{D\pi}(q^2)$ computed at its nearest pole $m_0^2=m_{D^\ast}^2$. In other words, and as indicated in eq.~(\ref{eq:s2}), we need to evaluate
\bea\label{res1}
 \displaystyle{\underset{q^2=m_{D^\ast}^2}{\rm Res}} f_+(q^2)= {1\over 2} m_{D^\ast} f_{D^\ast} g_{D^\ast D\pi}\,.
\eea 
\subsection{Pionic couplings $g_{D^\ast D\pi} \propto g_{c}$ and  $g_{B^\ast B\pi} \propto g_{b}$ }
Determination of the pionic coupling $g_{D^\ast D\pi}$ attracted quite a lot of interest in the literature. In the early works the QCD based estimate of that quantity relied on the method of QCD sum rules. The results, however, turned out to be very small and incompatible with a general picture that can be established when considering such couplings to a soft pion in the case of either mesons or baryons. In ref.~\cite{walain}  it was indeed claimed that the value of that coupling should be considerably larger (by at least a factor of two with respect to the QCD sum rule results), which was later confirmed by the first experimental measurement of the width of the charged vector meson, $\Gamma(D^{\ast \pm})$, made by CLEO~\cite{CLEO-Dstar}. The first lattice QCD estimate 
of this coupling was made in ref.~\cite{wgregorio} but in the quenched approximation and at two lattice spacings. The lattice results  clearly showed that the value was much larger than the one suggested by the results obtained by using different forms of QCD sum rules. Later on, two unquenched computations of that coupling have been performed at one or two lattice spacings and the results were reported in refs.~\cite{whaas,altug}. Finally a detailed unquenched analysis at four different fine lattice spacings and with several values of the sea quark mass, allowed to take the continuum limit, and the resulting value was~\cite{wfrancesco} 
\bea
g_{D^\ast D\pi}=16.0^{+1.1}_{-1.5} \,.
\eea
This coupling is often expressed in terms of $g_c$, defined via
\bea
g_{D^\ast D\pi} ={2\sqrt{m_Dm_{D^\ast}}\over f_\pi} \ g_c\,,
\eea
giving $g_c=0.54(3)(^{+2}_{-4})$. A redefinition of $g_{D^\ast D\pi}$ to $g_c$ is convenient because the latter is expected to scale as a constant with the heavy quark mass (up to power corrections $\propto 1/m_h^n$, with $m_h$ being the heavy quark mass).  
That value obtained on the lattice~\cite{wfrancesco} was then confirmed by the BaBar collaboration~\cite{gc-BaBar}, 
\bea\label{g-babar}
g_c=0.570(6),\quad g_{D^\ast D\pi} =16.92(19)\,,
\eea
which is very accurate and it is very unlikely that a lattice QCD computation of that coupling could be improved in the near future. 
In the case of $B$-mesons this coupling cannot be studied experimentally because the $B^\ast \to B\pi$ decay is kinematically forbidden and therefore one needs to rely on lattice QCD. Very recently a preliminary result for $g_b$, obtained from the analysis at two different lattice spacings, has been reported in ref.~\cite{g-ukqcd},
\bea\label{gb}
g_b= 0.569(76) \;\Rightarrow\; g_{B^\ast B\pi} =46(6)\,,
\eea
suggesting that the power corrections to $g_h$ are indeed very small ($h$ being a generic heavy quark mass). Indeed the values for $g_c$ and $g_b$ are perfectly consistent with each other and compatible with the values obtained in the static limit~\cite{g-static}.

With the value for 
$g_{D^\ast D\pi}$ ($g_c$) given in eq.~(\ref{g-babar}), the remaining information needed to evaluate the residuum~(\ref{res1}) is the vector meson decay constant $f_{D^\ast}$. In the following we present the results for $f_{D^\ast}$ obtained by means of numerical simulations of QCD on the lattice. Since we will also discuss the $B\to \pi\ell\nu_\ell$ decay, in the next subsection we will provide the first unquenched lattice QCD determination of $f_{B^\ast}$ and therefore, together with~(\ref{gb}), we will also be able to evaluate the residuum of the  $f_+^{B\pi}(q^2)$ form factor at its nearest pole $m_{B^\ast}^2$.   

\subsection{\label{sec-2}Determination of the vector meson decay constants on the lattice}

In this section we present our results for the decay constants of the heavy-light mesons, both the pseudoscalar and the vector ones. Our aim is to compute the ratio of the two and obtain a lattice QCD estimate of $f_{D^\ast}/f_D$ and $f_{B^\ast}/f_B$. The benefit of considering the ratios is that the uncertainties due to chiral extrapolations cancel out, and also a conversion of the results from lattice units to physical units does not become an issue when discussing the sources of systematic uncertainty in the final results. Furthermore, in the limit of the infinite heavy quark mass the pseudoscalar and vector heavy-light mesons are degenerate and the ratio of their decay constants is equal to one up to small perturbative corrections. That latter feature is particularly appealing for the lattice QCD computations in which the heavy quark is treated as fully propagating, i.e. without recourse to an effective theory of heavy quark on the lattice. In that situation the problem of reaching the physical $b$-quark mass through an extrapolation in inverse heavy quark mass, is circumvented and instead of extrapolating one actually interpolates the results to a desired point corresponding to $1/m_b$.  

We now briefly remind the reader of the method used to extract the heavy-light meson decay constants on the lattice. For that we need to compute the following correlation functions, 
\bea
\label{r1}
&& C_{P}(t)  =  \langle {\displaystyle \sum_{\vec x} }   P(\vec x; t) P^\dagger(0; 0) \rangle = - (\mu_h+m_q)^2\sum_{\vec x} \langle {\rm Tr}\left[ S_h(\vec 0,0;\vec x,t)\gamma_5 S_q (\vec x,t;\vec 0,0) \gamma_5 \right] \rangle \nn\\
&&\qquad\qquad\qquad \xrightarrow[]{\displaystyle{ t\gg 0}} \; \frac{\cosh[  m_{H_q} (T/2-t)]}{ m_{H_q} }  \left| \langle 0\vert P(0)
\vert H_q (\vec 0) \rangle \right|^2 e^{- m_{H_q} T/2},\nn
\eea
\bea\label{r1bis}
&& C_{V}(t)  =  \frac{1}{3}\langle {\displaystyle \sum_{i, \vec x} }   V_{i}(\vec x; t) V^\dagger_{i}(0; 0) \rangle = - \frac{1}{3} Z_A^2\sum_{i,\vec x} \langle {\rm Tr}\left[ S_h(\vec 0,0;\vec x,t)\gamma_i S_q (\vec x,t;\vec 0,0) \gamma_i\right] \rangle \nn\\
&&\qquad \xrightarrow[]{\displaystyle{ t\gg 0}} \; \frac{\cosh[  m_{H_q^\ast} (T/2-t)]}{ 3 m_{H_q^\ast} }  \sum_{i} \left| \langle 0\vert V_i(0)
\vert H_q^\ast (\vec 0, \lambda) \rangle \right|^2 e^{- m_{H_q^\ast} T/2}.
\eea
In the above expression the interpolating operator used to discuss the properties of the heavy-light pseudoscalar meson is $P=(\mu_h+\mu_q) \bar h \gamma_5 q$, where $\mu_{h(q)}$ 
stands for a heavy (light) bare quark mass. This choice is particularly appealing when working with twisted mass QCD on the lattice because it is renormalization  group invariant and 
therefore no renormalization constant is needed to compute the corresponding meson decay constant $f_{H_q}$. As for the vector meson on the lattice we use 
$V_i = \bar h \gamma_i q$, with $i=1,2,3$. For the computation of the vector meson decay constant $f_{H_q^\ast}$, we need the renormalization constant $Z_A(g_0^2)$ which was already computed non-perturbatively in ref.~\cite{ZZZ}.  
In eq.~(\ref{r1}) we also wrote the spectral decomposition of the correlation function in terms of hadronic states, dominated by the lowest lying one for large time separations between the interpolating field operators. In these expressions we accounted for the fact that the periodicity of our lattice allows us to average $[C_{P,V}(t)+C_{P,V}(T-t)]/2$, where $T$ is the size of our lattice in time direction. Notice also that in the above notation the light quark propagator is denoted by $S_q(\vec x,t; \vec 0, 0) = \langle \bar q(x)q(0)\rangle$, and similarly for the heavy quark one. Finally the decay constants are related to the hadronic matrix elements as,
\bea\label{r2}
&& \langle 0\vert P \vert H_q (\vec 0) \rangle =   f_{H_q} m_{H_q}^2 \,,\nn\\
&&\hfill\nn\\
&& \langle 0\vert V_i \vert  H_q^\ast (\vec 0,\lambda)  \rangle = f_{H_q^\ast} m_{H_q^\ast}  e_i^\lambda \,, 
\eea
where $e_\mu^\lambda$ is the polarization vector of the vector heavy-light meson $H^\ast_q$. 
In eq.~(\ref{r1}) we assumed the source operators to be local, as needed for the extraction of decay constants. To make sure the lowest lying state is well isolated we also combine the local operators with those obtained by implementing the Gaussian smearing procedure. The smearing parameters we choose are the same as those already discussed in refs.~\cite{charm2,wantoine}.
\begin{table}[h!!]
\centering 
{\scalebox{.93}{\begin{tabular}{|c|cccccc|}  \hline \hline
{\phantom{\huge{l}}}\raisebox{-.2cm}{\phantom{\Huge{j}}}
$ \beta$& 3.8 &  3.9  &  3.9 & 4.05 & 4.2  & 4.2    \\ 
{\phantom{\huge{l}}}\raisebox{-.2cm}{\phantom{\Huge{j}}}
$ L^3 \times T $&  $24^3 \times 48$ & $24^3 \times 48$  & $32^3 \times 64$ & $32^3 \times 64$& $32^3 \times 64$  & $48^3 \times 96$  \\ 
{\phantom{\huge{l}}}\raisebox{-.2cm}{\phantom{\Huge{j}}}
$ \#\ {\rm 
 meas.}$& 240 &  240  & 240  & 750/404 & 288 & 480  \\ \hline 
{\phantom{\huge{l}}}\raisebox{-.2cm}{\phantom{\Huge{j}}}
$\mu_{\rm sea 1}$& 0.0080 & 0.0040 & 0.0030 & 0.0030 & 0.0065 &  0.0020   \\ 
{\phantom{\huge{l}}}\raisebox{-.2cm}{\phantom{\Huge{j}}}
$\mu_{\rm sea 2}$& 0.0110 & 0.0064 & 0.0040 & 0.0060 &   &     \\ 
{\phantom{\huge{l}}}\raisebox{-.2cm}{\phantom{\Huge{j}}}
$\mu_{\rm sea 3}$&  & 0.0085 &  & 0.0080 &   &     \\ 
{\phantom{\huge{l}}}\raisebox{-.2cm}{\phantom{\Huge{j}}}
$\mu_{\rm sea 4}$&  & 0.0100 &  &   &   &     \\   \hline 
{\phantom{\huge{l}}}\raisebox{-.2cm}{\phantom{\Huge{j}}}
$a \ {\rm [fm]}$&   0.098(3) & 0.085(3) & 0.085(3) & 0.067(2) & 0.054(1) & 0.054(1)      \\ 
{\phantom{\huge{l}}}\raisebox{-.2cm}{\phantom{\Huge{j}}}
$Z_A (g_0^2)$~\cite{ZZZ}& 0.746(11) & 0.746(6) & 0.746(6)  & 0.772(6) & 0.780(6) & 0.780(6) \\ 
{\phantom{\huge{l}}}\raisebox{-.2cm}{\phantom{\Huge{j}}}
$\mu_{c}$~\cite{Blossier:2010cr}& 0.2331(82)  &0.2150(75)  &0.2150(75)   & 0.1849(65) & 0.1566(55) & 0.1566(55)  \\ 
 \hline \hline
\end{tabular}}}
{\caption{\footnotesize  Summary of the lattice ensembles used in this work (more details can be found in ref.~\cite{boucaud}).  
Data obtained at different $\beta$'s in this work are rescaled by using $r_0/a$, and the overall lattice spacing is fixed by matching $f_\pi$ obtained on the lattices with its physical value, leading to  $r_0= 0.440(12)$~fm (c.f. ref.~\cite{Blossier:2010cr}). All quark masses are given in lattice units.  }}
 \label{tab:0}
 \end{table}

\begin{figure}[t!!]
\begin{center}
\includegraphics[scale=.96]{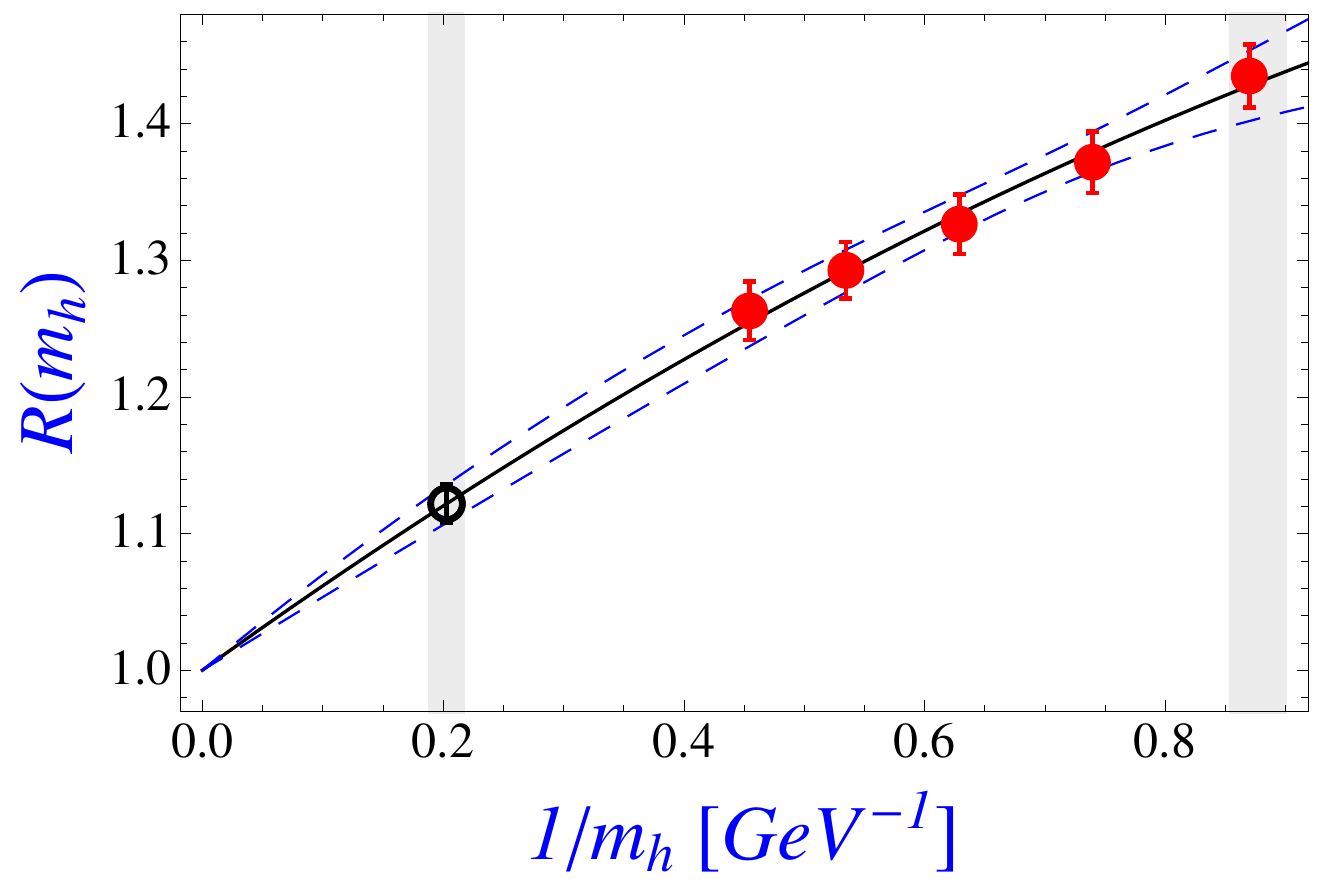}
\caption{ {\footnotesize 
Interpolation of the ratio $R(m_h)$, defined in eq.~(\ref{eq:RR}), using the expression~(\ref{eq:interp}) to the inverse $b$-mass point (empty symbol in the plot). The gray vertical stripes correspond to the $1/m_b$  and $1/m_c$ respectively. Quark mass definition used in this work is $m_h\equiv m_h^\msbar (2\ \gev)$. }}
\label{fig:2}
\end{center}
\end{figure}

In this work we use the gauge field configurations generated by the European Twisted Mass Collaboration (ETMC), in which the effect of $\nf=2$ dynamical (``{\sl sea}") light quarks has been included by using the Wilson regularization of QCD on the lattice with the maximally twisted mass term~\cite{fr}. Main information about the lattices used in this work are given in tab.~\ref{tab:0}. With respect to what has been presented in ref.~\cite{wcecilia} the results we present here are obtained with significantly larger statistics and by implementing the smearing procedure to make sure the signals for the lowest lying states are better isolated. In addition, in this paper we make an extra step and compute the decay constants for a series of heavy quark masses that will eventually help us interpolating to the physically interesting case, i.e. the one corresponding to the inverse $b$ quark mass. 
The results obtained by fitting the correlation functions to the form indicated in eq.~(\ref{r1}) and by using the definitions given in eq.~(\ref{r2}) are presented in Appendix~\ref{app1}. 
We then consider the ratio of decay constants
\bea
F(m_h,m_q) = {f_{H^\ast_q}\over f_{H_q}}\,,
\eea
where we specify the valence quark content, heavy quark mass $m_h$ and the mass of the light one $m_q$. The latter has been fixed to be equal to the mass of the light sea quark, i.e. the mass of the quark generated dynamically by using Hybrid Monte Carlo in the generation of the gauge field configurations. As for the heavy quark mass, in our first step we compute the above ratio with the bare heavy quark corresponding to that of the physical charm quark ($\mu_h=m_c$) the value of which has been fixed in ref.~\cite{Blossier:2010cr}. 
We then proceed and make a combined continuum and chiral extrapolation of our data as
\bea\label{eq:contR}
F(m_h,m_q)  = b^{(0)}_h +  b^{(1)}_{h} {m_{q}\over m_s} + b^{(2)}_{h} {a^2\over (0.086\ {\rm fm})^2} \,.
\eea
where in that extrapolation we use the ratio of the sea quark mass with respect to the physical strange quark mass (already computed in ref.~\cite{Blossier:2010cr} by using the same ensembles of the gauge field configurations) which then makes the parameter $b^{(1)}_{h}$ dimensionless. Similarly, and in order to make the value of the coefficient $b^{(2)}_{h}$ more informative 
in terms of smoothness of the continuum extrapolation, we divided the lattice spacing $a$ by its value at $\beta=3.90$. 
Note however that the leading discretization effects for the quantities computed by using twisted mass lattice QCD are of the order of $a^2$. 
\begin{figure}[t!!]
\begin{center}
\includegraphics[scale=.92]{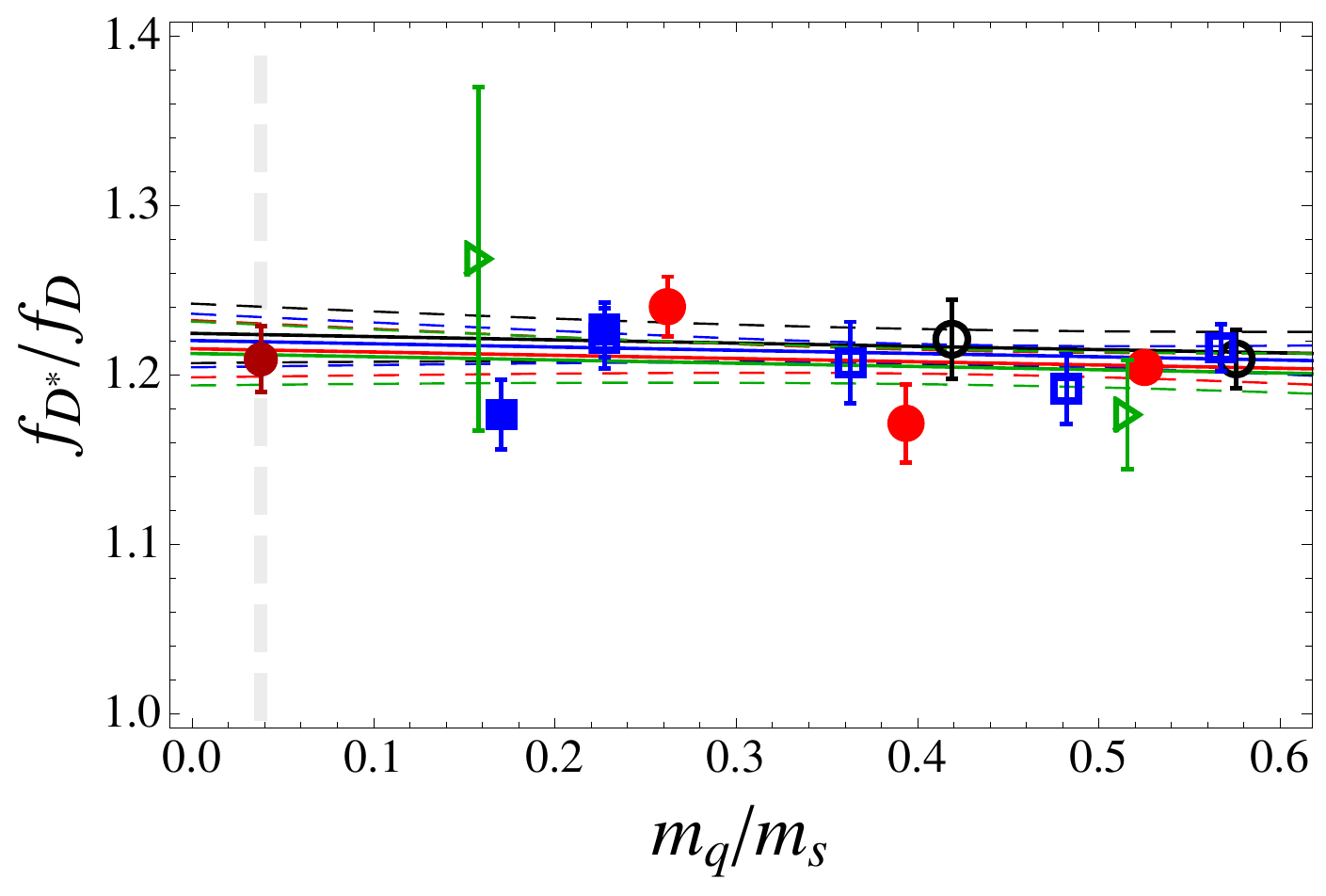}
\caption{{\footnotesize 
Chiral and continuum extrapolation of $f_{D^\ast}/f_D$, according to eq.~(\ref{eq:contR}) at four lattice spacings. 
The symbols corresponding to the lattice data are: {\Large\color{Black}\bf{\textopenbullet}} for $\beta=3.80$, $\boldsymbol{\color{blue} \boldsymbol{\square}}$ for $\beta=3.90$ ($L=24$), 
${\color{blue} \large \blacksquare}$ for $\beta=3.90$ ($L=32$), 
$ {\color{red} \mathlarger{\mathlarger{\mathlarger{\mathlarger{\bullet}}}}} $ for $\beta=4.05$,  and  ${\color{OliveGreen} \mathlarger{\mathlarger{\mathlarger{\triangleright}}}}$ for $\beta=4.20$. 
Full lines, corresponding to the fits at fixed lattice spacing, are indistinguishable due to the fact that the discretization 
effects on the ratio seem to be negligible (dashed lines are the upper/lower fit curves at fixed lattice spacing). The vertical gray line corresponds to the physical 
$m_{u,d}/m_s$ point.  
 }}
\label{fig:1}
\end{center}
\end{figure}
The results of that extrapolation expressed in terms of coefficients $b^{(0,1,2)}_{h}$,  for each of our heavy quark masses are presented in tab.~\ref{tab:1}, 
and for the case of the $D^{(\ast)}$-meson decay constants illustration is provided in fig.~\ref{fig:1}. We notice that the extrapolations are smooth and 
the discretization effects are indeed small for not so large values of the heavy quark mass. In particular, by using $F(m_c,m_{ud})^{\rm cont.}\equiv {f_{D^\ast}/f_D}= b^{(0)}_c +  b^{(1)}_{c}\times ({m_{ud}/m_s})$, and $m_{ud}/m_s=0.037(1)$~\cite{Blossier:2010cr}, we obtain 
\bea\label{eq:cc}
{ f_{D^\ast}\over f_D} = 1.209\pm 0.020\,.
\eea
\begin{table}[t]
\centering 
{\scalebox{.93}{\begin{tabular}{|c|ccc|}  \hline 
{\phantom{\huge{l}}}\raisebox{-.2cm}{\phantom{\Huge{j}}}
$n\qquad$& $b^{(0)}_h$ &  $b^{(1)}_{h}$  &  $b^{(2)}_{h}$    \\  \hline
{\phantom{\huge{l}}}\raisebox{-.2cm}{\phantom{\Huge{j}}}
$0\qquad$  &  $1.208\pm 0.024$  &    $0.011\pm 0.016$   & $-0.016\pm 0.028$  \\ 
{\phantom{\huge{l}}}\raisebox{-.2cm}{\phantom{\Huge{j}}}
$1\qquad$  &  $1.186\pm 0.024$  &    $0.025\pm 0.017$   & $-0.018\pm 0.029$  \\ 
{\phantom{\huge{l}}}\raisebox{-.2cm}{\phantom{\Huge{j}}}
$2\qquad$  &  $1.167\pm 0.025$  &    $0.042\pm 0.018$   & $-0.022\pm 0.030$  \\ 
{\phantom{\huge{l}}}\raisebox{-.2cm}{\phantom{\Huge{j}}}
$3\qquad$  &  $1.155\pm 0.026$  &    $0.060\pm 0.020$   & $-0.029\pm 0.030$  \\ 
{\phantom{\huge{l}}}\raisebox{-.2cm}{\phantom{\Huge{j}}}
$4\qquad$  &  $1.141\pm 0.028$  &    $0.079\pm 0.023$   & $-0.022\pm 0.033$  \\ 
{\phantom{\huge{l}}}\raisebox{-.2cm}{\phantom{\Huge{j}}}
$5\qquad$  &  $1.139\pm 0.031$  &    $0.107\pm 0.026$   & $-0.038\pm 0.040$  \\ 
 \hline
\end{tabular}}}
{\caption{\footnotesize  Results of the chiral and continuum extrapolation of our data for each of the heavy quark masses $m_h = \lambda^n m_c$ obtained by using eq.~(\ref{eq:contR}). 
}}
 \label{tab:1}
 \end{table}
To that one needs to add the uncertainty arising from the error on the charm quark mass, that we discuss further below. The ratio $F(m_h,m_{ud})$ is equal to one in the limit of infinite heavy quark mass up to calculable 
perturbative corrections stemming from the matching between the full QCD (in which we compute our decay constants) and the heavy quark effective theory (in which this scaling law is verified by construction). The matching coefficient 
has been computed to next-to-next-to leading order and reads~\cite{grozin},
\bea\label{eq:match}
C(\mu) = 1 - {2\over 3} {\alpha_s(\mu)\over \pi} + \left[ -\frac{1}{9} \zeta(3) +\frac{2}{27}\pi^2 \log 2 +\frac{4}{81}\pi^2 +\frac{115}{36} \right]\left( {\alpha_s(\mu)\over \pi}\right)^2\,.
\eea
In our next step we construct the quantity 
\bea\label{eq:RR}
R(m_h) = {\, F(m_h,m_{ud})^{\rm cont.}\over C(m_h)}\,,
\eea
which indeed has a property that 
\bea\label{eq:static}
\lim_{m_h\to\infty} R(m_h,m_{ud})=1\,.
\eea
Finally, we can interpolate in the heavy quark mass by using  
\bea\label{eq:interp}
{R(m_h)}= 1 +{ \alpha_1\over m_h}+ { \alpha_2\over m_h^2} + \dots
\eea
where $\alpha_{1,2}$ are the coefficients obtained from the fit. 
The result of that interpolation in the case of $m_h=m_b$ then leads to the desired $f_{B^\ast}/f_B= C(m_b) R(m_b)$. 
In practice, we compute 
compute the ratio $R(m_h)$ for a series of heavy quark masses successively increased by a factor of $\lambda$, namely $m_h\equiv m_n=\lambda^n m_c$.  
We choose $\lambda = 1.175$. Clearly, for larger heavy quark masses, closer to the $b$-quark mass, the terms proportional to higher powers in lattice spacing become 
important and the results of extrapolation obtained by using eq.~(\ref{eq:contR}) become less reliable. Furthermore we observe that the statistical errors of our 
data progressively increase with increase of the heavy quark mass. Knowing however that the final result is obtained through an interpolation~(\ref{eq:interp}) the impact of the inclusion of 
 results obtained with larger heavy quark masses (close to the physical $b$-quark mass) becomes essentially irrelevant because of the heavy quark symmetry formula~(\ref{eq:static}).
The result obtained by using $3$, $4$, or $5$ points in interpolation are actually fully compatible. As a final we quote the value obtained by using $n=4$, i.e. the heavy 
charm quark mass and four successively larger masses $m_n=\lambda^n m_c$. We obtain
\bea\label{eq:bb}
{ f_{B^\ast}\over f_B} = 1.050\pm 0.013\,,
\eea
and for the parameters in eq.~(\ref{eq:interp}) we find,
\bea
\alpha_1=0.633(74)~\gev\,,\qquad \alpha_2=-0.16(10)~\gev^2\,.
\eea
The result of that interpolation is also shown in fig.~\ref{fig:2}.
Since we have the values of $\alpha_{1,2}$ needed in eq.~(\ref{eq:interp}) we can now estimate the error to the ratio of decay constants coming from uncertainty on the physical charm and $b$-quark masses the value of which are~\cite{Blossier:2010cr,Dimopoulos:2011gx}, namely
\bea
m_c^\msbar(2\ \gev)=1.14(4)~\gev\,,\qquad m_b^\msbar(2\ \gev)=4.93(16)~\gev\,.
\eea
That variation has been depicted by the thick vertical lines in fig.~\ref{fig:2} corresponding to $\delta (1/m_b)$ and $\delta (1/m_c)$ respectively. Inclusion of these variations in the uncertainty of our ratios of decay constants gives
\bea
{ f_{D^\ast}\over f_D} = 1.208\pm 0.027\,,\quad { f_{B^\ast}\over f_B} = 1.050\pm 0.016\,,
\eea
which is our final result for $f_{D^\ast}/f_D$. We note that our result for $f_{B^\ast}/f_B$ remains stable regardless of the fact that we include in interpolation $3$, $4$, $5$ or $6$ points (the error on the result of interpolation is unchanged after $4$ points included).
Another source of uncertainty for $f_{B^\ast}/f_B$ comes from the use of $\alpha_s(m_h)$ in eqs.~(\ref{eq:match},\ref{eq:RR}). The above result~(\ref{eq:bb}) is obtained by using the physical running coupling corresponding to $\alpha_s(M_Z)=0.1176$. However, by using the running coupling in the theory with $N_{\rm f}=2$ flavors, i.e. with $\Lambda_{N_{\rm f}=2}^\msbar = 330^{+21}_{-54}$~MeV~\cite{FLAG,lambda2}, our result remains within the error bars obtained before. Finally,  one may worry about the higher perturbative corrections in the matching constant~(\ref{eq:match}) which were discussed in ref.~\cite{grozin}. Inclusion of the three-loop coefficient gives the result fully consistent with the one obtained above. To be more specific:
\bea
 { f_{B^\ast}\over f_B} = 1.050\pm 0.016   \left. {}^{+0.006}_{-0.000}\right|_{\alpha_s}   \left. {}^{+0.000}_{-0.003}\right|_{C^{\rm NNNLO}}   \left. {}^{+0.005}_{-0.004}\right|_{\# \ {\rm points}}\,,
 \eea
where the last error reflects the effect of the number of points used in the interpolation to $1/m_b$. The central value is obtained with $5$ points and the change in central value if we use $4$ or $6$ points is reflected in that last error. Combining the errors in quadrature we finally obtain our final result for $f_{B^\ast}/f_B$, namely 
\bea\label{eq:FINAL}
{ f_{D^\ast}\over f_D} = 1.208\pm 0.027\,,\quad  { f_{B^\ast}\over f_B} =1.051\pm 0.017\,.
 \eea
Two comments are in order. With respect to our previous estimate of this ratio in the case of $D$ mesons, the central value is lower but the results is of course consistent with the published one~\cite{wcecilia} within the error bars (the new one being more accurate). As for the ratio $f_{B^\ast}/ f_B$ there are only a few published lattice results. Two quenched computation with Wilson fermions lead to 
$f_{B^\ast}/f_B=1.07(5)$~\cite{wfred}, and $1.06(3)$~\cite{UKQCD-fB}. Also quenched, but a result obtained by using an effective theory treatment of the heavy quark on the lattice has been obtained in ref.~\cite{MILC-old} where also a continuum extrapolation has been implemented to get $f_{B^\ast}/f_B=1.01(1)(^{+4}_{-1})$. Our result is obviously a significant improvement with respect to these previous lattice QCD estimates. 
Finally, we should mention that very recently a computation of these ratios has been made in the framework of the QCD sum rules. More specifically,  the results of two independent studies are:
$f_{D^\ast}/f_D= 1.20(^{+10}_{-07})$  and  $f_{B^\ast}/f_B= 1.02(^{+7}_{-3})$ in ref.~\cite{khodjV}, and $f_{D^\ast}/f_D= 1.209(22)$  and  $f_{B^\ast}/f_B= 1.031(8)$ in ref.~\cite{narisonV}, are in very good agreement with our findings~(\ref{eq:FINAL}). The authors of ref.~\cite{MSS} reported another QCD sum rule result, $f_{D^\ast}/f_D= 1.22(8)$, also consistent with our result.

\subsection{The residuum value}

With the above ingredients we can now compute the residuum. To do so we will use the experimentally established value for the $D$-meson decay constant $f_D=204(5)$~MeV~\cite{PDG}, and we finally obtain
\bea\label{eq:resD}
 \displaystyle{\underset{q^2=m_{D^\ast}^2}{\rm Res}} f^{D\pi}_+(q^2) = {1\over 2} m_{D^\ast} {f_{D^\ast}\over f_D} f_D\  g_{D^\ast D\pi} =4.19(15)~\gev^2,
 \eea
where for the coupling $g_{D^\ast D\pi}$ we used the value given in eq.~(\ref{g-babar}). 
Similarly, by using $g_{B^\ast B\pi}$ from eq.~(\ref{gb}) and by averaging the unquenched lattice QCD determinations of the $B$-meson decay constant $f_{B}=188(6)$~MeV~\cite{FLAG,fB-latt}, we get
\bea\label{eq:resB}
 \displaystyle{\underset{q^2=m_{B^\ast}^2}{\rm Res}}  f^{B\pi}_+(q^2) = {1\over 2} m_{B^\ast} {f_{B^\ast}\over f_B} f_B\  g_{B^\ast B\pi} =24.3(3.4)~\gev^2. 
 \eea

\section{Constraining the second pole residuum\label{sec-res2}}

Next pole contributing the $f_+^{D\pi}(q^2)$ form factor is the radially excited vector meson $D^{\ast \prime}$, for which we will take the value obtained in ref.~\cite{delAmoSanchez:2010vq},  $m_{D^{\ast \prime}}=2609(4)$~MeV and $\Gamma=93(14)$~MeV.
By using
\bea
\Gamma(D^{\ast \prime +}\to D^0 \pi^+) = {g_{D^{*\prime}D^0\pi^+}^2 \over 24 \pi m_{D^{\ast \prime}}^2} |\vec k_\pi |^3,
\eea
where $2 m_{D^{\ast \prime }}  |\vec k_\pi | = \lambda^{1/2}(m_{D^{\ast \prime }}^2, m_{D}^2,m_{\pi}^2)$, and assuming the width to be entirely saturated by $D^{(\ast)}\pi$, we obtain 
\bea
\Gamma(D^{\ast \prime 0}) &=& 
\Gamma(D^{\ast \prime 0}\to D \pi ) +\Gamma(D^{\ast \prime 0}\to D^\ast \pi) =\left( 1+ {1\over r_{DD^\ast}}\right)\Gamma(D^{\ast \prime 0}\to D \pi ) \nn\\
&=& \left( 1+ {1\over r_{DD^\ast}}\right)\ \left[ \Gamma(D^{\ast \prime 0}\to D^- \pi^+) +\Gamma(D^{\ast \prime 0}\to D^- \pi^0) \right]\nn\\
&=& \left( 1+ {1\over r_{DD^\ast}}\right)\ {g_{D^{*\prime 0}D^-\pi^+}^2  \over 16 \pi m_{D^{\ast \prime}}^2}  |\vec k_{\pi} |^3 \,,
\eea
where we used the experimentally established ratio, $ r_{DD^\ast}=\Gamma(D^{\ast \prime 0}\to D \pi )/\Gamma(D^{\ast \prime 0}\to D^\ast \pi ) = 0.32(9)$~\cite{delAmoSanchez:2010vq}, and in the last line the isospin symmetry.~\footnote{The heavy quark symmetry argument suggests the factor $ r_{DD^\ast}$ to be close to one. The fact that it is very different from one could be explained in quark models by the fact that the wave function of the radial excitation contains a node the position of which can have significant consequences on the symmetry relations. Nevertheless, the main conclusions of this paper remains unchanged even if we adopt the value $r_{DD^\ast}\approx 1$.}  The number we obtain is 
$|g_{D^{*\prime}D\pi^+} |= 5.6(7)$  for which we do not {\it a priori} know the sign.~\footnote{Clearly the value of  $|g_{D^{*\prime}D\pi^+}|$ 
obtained with this assumption is an overestimate as the other multi-pion channels open up, each giving a positive definite contribution to the width of $D^{\ast \prime}$.  } 
To get the residuum,
\beq\label{ref:s4}
 \displaystyle{\underset{q^2=m_{D^{\ast\prime}}^2}{\rm Res}} f^{D\pi}_+(q^2)
 =\frac{1}{2}m_{D^{*\prime}}\,{f_{D^{*\prime}}\over f_{D^\ast}}\, {f_{D^\ast}\over f_D} \, f_D\, g_{D^{*\prime}D^0\pi^+}\,
\eeq
we also need  $f_{D^{*\prime}}$. For that we can rely on the recent lattice QCD estimate, $f_{D^\prime}/f_D= 0.57(16)$, made in ref.~\cite{wantoine} 
and assume that $f_{D^{\ast \prime}}/f_{D^\ast}= f_{D^\prime}/f_D$, which then gives us  $f_{D^{*\prime}}=140(40)$~MeV. That result fairly agrees with a very recent evaluation of the same decay constant by using the QCD sum rules, $f_{D^{*\prime}}=183(^{+13}_{-24})$~MeV~\cite{Gelhausen:2014jea}.

With the above results we can have an estimate of the residue~(\ref{ref:s4}) and obtain,
\beq \label{eq:res4} 
 \displaystyle{\underset{q^2=m_{D^{\ast\prime}}^2}{\rm Res}} f^{D\pi}_+(q^2)= -(1.0 \pm 0.3)~\gev^2.
\eeq
Had we used the results recently obtained  by LHCb, $m_{D^{\ast\prime }} =2.649(5)$~GeV and $\Gamma(D^{\ast\prime })=140(25)$~MeV~\cite{LHCb3}, the value of the residuum would become $-1.2(4)~\gev^2$, fully compatible with eq.~(\ref{eq:res4}).

\underline{The question of the sign of residua}. This question deserves careful discussion because it is essential in the fits which are performed later. We attributed the minus sign to the residue in eq.~(\ref{eq:res4}), with respect to the ground state one which is positive by convention -- just like the form factor itself. In fact the sign in eq.~(\ref{eq:res4}) can be predicted by theory only if the strong coupling ($g_{D^{*\prime}D\pi}$) and the decay constant ($f_{D^{*\prime}}$) were computed simultaneously and also compared to the residuum of the form factor at the first pole. This is precisely what is provided by the lattice QCD calculation of the matrix element of $A_\mu = \bar u \gamma_\mu\gamma_5 d$. Such a lattice computation has been performed in ref.~\cite{Blossier:2013qma} for the matrix element between the first radial excitation and the ground state of heavy-light mesons in the static limit. The computation was made for the spatial component $\langle H|A_i |H^{\ast\prime}\rangle$, at the zero three-momentum transfer $|\vec q|=0$. Unhappily this is not what one needs for the present purpose. In particular, to get $g_{D^{*\prime}D\pi} $ through PCAC (partial conservation of the axial current), one  needs the matrix element at $q^2=m_\pi^2\approx 0$, i.e. $|\vec q|=\Delta E \equiv m_1-m_0$ (splitting between the first radial excitation and the ground state). 
Moreover, one  also needs to compute the time component $A_0$ because 
\bea
g_{D^*_nD\pi} = - {2\sqrt{ m_{D_n^\ast} m_D}\over f_\pi} {\, q^\mu\, \over |\vec q| } \biggl. \langle D_n^\ast \vert \bar u \gamma_\mu\gamma_5 d\vert D\rangle\biggr|_{q^2= 0}.
\eea 
Note that the missing matrix element  $\langle H|A_0|H^{\ast\prime}\rangle$ is non-zero and $q^0\neq 0$ as well. 
In principle the limit $q^2= 0$ could be taken by using the lattice methodology developed in ref.~\cite{wemmanuel} or by using a quark model which reliably reproduces radial distributions of the axial charge in the heavy-light systems. 
In ref.~\cite{wemmanuel} two such models were shown to be reliable, namely the one based on the Dirac equation (with a simple Coulombic plus a linear confining potential) and those belonging to a class known as the Bakamjian-Thomas models. 
Since the relevant lattice computation has not yet been done, we can presently recourse only to quark models for the theoretical estimate of the coupling constant and for the sign of the residuum. 
We checked that the two quark models lead to  consistent conclusions, namely that (i) the matrix element of the spatial component of the axial current diminishes rapidly when passing from $|\vec q| =0$  to $|\vec q|  = \Delta E$ (note that $q^2=0$ implies $q^0 = |\vec q|  = \Delta E$), and can even change sign depending on the model, (ii) however, the temporal component of the axial current changes from zero to a large value which is much larger than the spatial contribution. In this way $\langle H|A_0|H^{\ast\prime}\rangle/|\vec q|$ remains rather stable and is quite similar in both models.
Therefore the desired overall sign is reliably obtained from the temporal component of the axial current if we also can compute the decay constant with the same convention. The latter can be done only within the Bakamjian-Thomas approach where one indeed verifies that the sign of the residuum $f_{D^{*\prime}}\,g_{D^{*\prime}D\pi}$ is opposite to the sign of $f_{D^{*}}\,g_{D^{*}D\pi}$. This is also consistent with observations based on the use of $^3P_0$ model, as discussed in ref.~\cite{wjerome}. Taking the decay constant to be positive, the negative sign is then attributed to  $g_{D^{*\prime}D\pi}$. 

We also use the two models to compute the absolute value of $g_{D^{*\prime}D\pi}$  ($g^{\prime}$). Extracted from the  width of $D^{\ast\prime}$ we have $g_{D^{*\prime}D \pi} =- 5.6(7)$. If instead we use the quark model to obtain a rough estimate of the pionic coupling in the $q^2 =0$ and in the static $m_h\to \infty$ limit, we obtain $g^\prime \in (0.1,0.2)$. 
As we saw in the case of $g_{D^{*}D\pi}$, the couplings $g_c$ and $g_b$ were fully compatible with each other and with the value computed in the static limit of QCD on the lattice. Assuming  the same situation holds true for the axial couplings involving the radial excitation and the ground state we would have for the quark model estimate
\bea\label{eq;gprime}
g_{D^{*\prime}D \pi} = { 2 \sqrt{  m_D m_{D^{\ast \prime} }}\over f_\pi} g^\prime = -(5.1\pm 1.7)_{\rm Quark \ Models} \,,
\eea
which is in quite good agreement with the above value extracted from the width of $D^{\ast\prime}$. In what follows, for the residuum of the form factor $f_{+}^{D\pi}(q^2)$ at the second pole we will take the number given in eq.~(\ref{eq:res4}).

A similar discussion can be extended to the case of $B$-mesons. We saw in the case of the couplings to the nearest pole that $g_c$ and $g_b$ were practically indistinguishable and compatible with the values found in the static heavy quark limit. Assuming the same holds true in the case of the coupling to the radially excited state, our $g_{B^{*\prime}B \pi}$ coupling can be obtained from 
\bea\label{DtoB}
g_{B^{*\prime}B \pi} = \sqrt{{m_{B^{*\prime}}m_B\over m_{D^{*\prime}}m_D} } \ g_{D^{*\prime}D \pi}\,.
\eea
By using $g_{D^{*\prime}D \pi} =- 5.6(7)$ obtained from the width of $D^{*\prime}$, and $m_{B^{\ast \prime}} = 5.97(1)$~GeV, isospin average of the recently observed states at Tevatron~\cite{Tevatron}, we get $g_{B^{*\prime}B \pi}=-14(2)$. 
As we saw, earlier in this section, the value for the decay constant of the radially excited $D^{(\ast)}$ obtained by the QCD sum rules in ref.~\cite{Gelhausen:2014jea} agrees rather well with the value obtained in lattice QCD. Since there is no lattice QCD estimate of $f_{B^{*\prime}}$, we will rely on the QCD sum rule value $f_{B^{*\prime}}=165(^{+46}_{-12})$~MeV, which finally leads to, 
\beq \label{eq:res5} 
\displaystyle{\underset{q^2=m_{B^{\ast\prime}}^2}{\rm Res}} f_{+}^{B\pi}(q^2)= -(7.7\pm 1.6)~\gev^2,
\eeq
where we combined in quadrature the errors coming from $g_{B^{*\prime}B \pi}$  and $f_{B^{*\prime}}$. 

\section{Superconvergence, heavy-quark scaling, and the effective pole model \label{super}}

Let us summarize our findings so far. 
We replaced the discontinuity in the dispersion relation for the form factor $f_+^{D\pi}(q^2)$ in eq.~(\ref{eq:unstr})  by the infinite series of (narrow states) poles, namely 
\bea\label{3poles}
f_+(q^2) ={ \gamma_0 \over  m_{D^\ast}^2 - q^2} + { \gamma_1 \over  m_{D^{\ast\prime}}^2 - q^2}+ { \gamma_2 \over  m_2^2 - q^2} + \dots \,,
\eea
where, for shortness, from now on we will denote the residua as $\gamma_n$. The value of $\gamma_0 = \displaystyle{\underset{q^2=m_{D^{\ast }}^2}{\rm Res}}f_{+}^{D\pi}(q^2)$ is by now very well known~(\ref{eq:resD}), while the one corresponding to the second pole, $\gamma_1= \displaystyle{\underset{q^2=m_{D^{\ast\prime}}^2}{\rm Res}}f_{+}^{D\pi}(q^2)$, is reasonably well estimated too~(\ref{eq:res4}). We also noted that the two residua have different sign.  The information about the higher singularities is more difficult to obtain. 

\begin{figure}[!htbp!] 
  \begin{center}
\includegraphics[height=7.5cm]{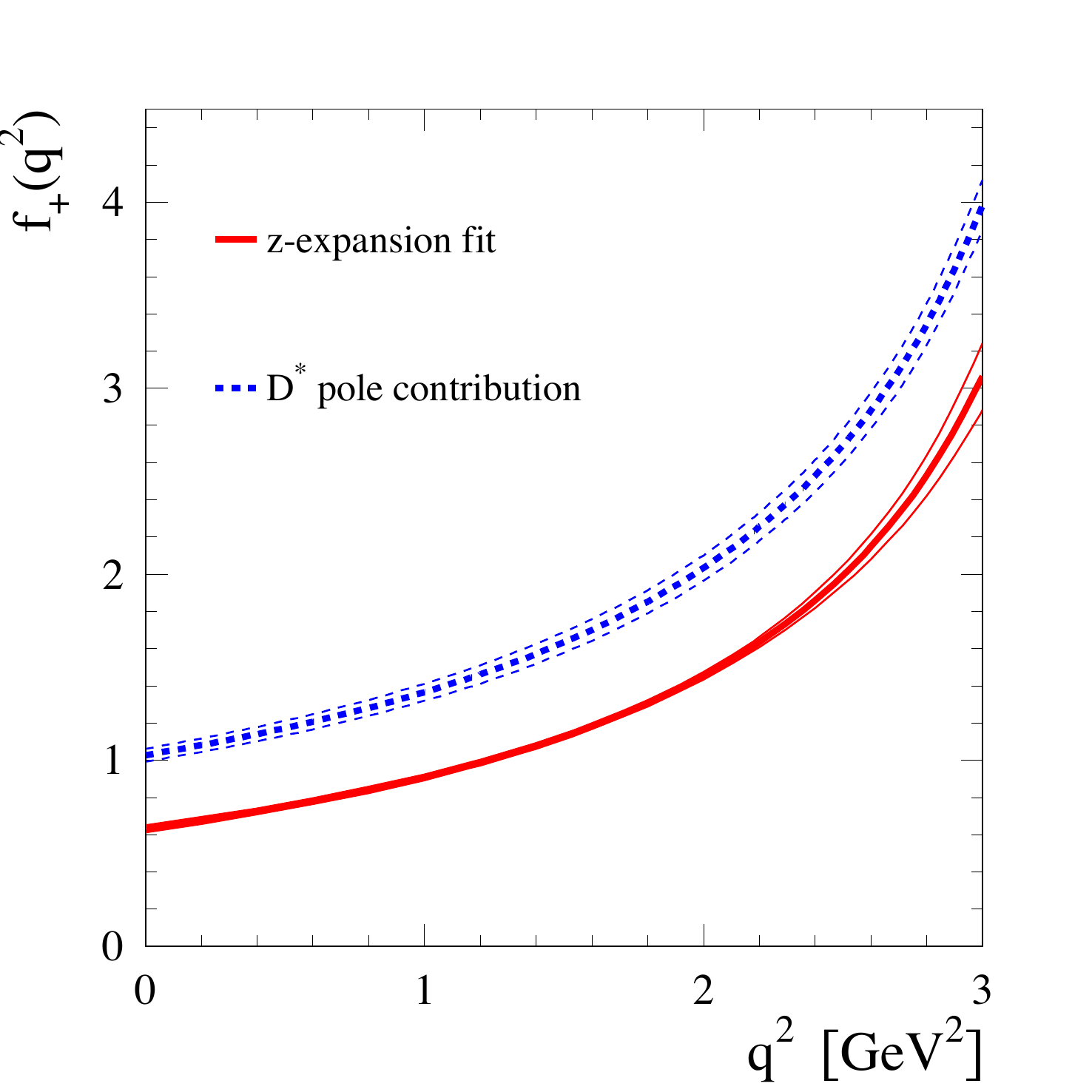}~\qquad~\includegraphics[height=7.5cm]{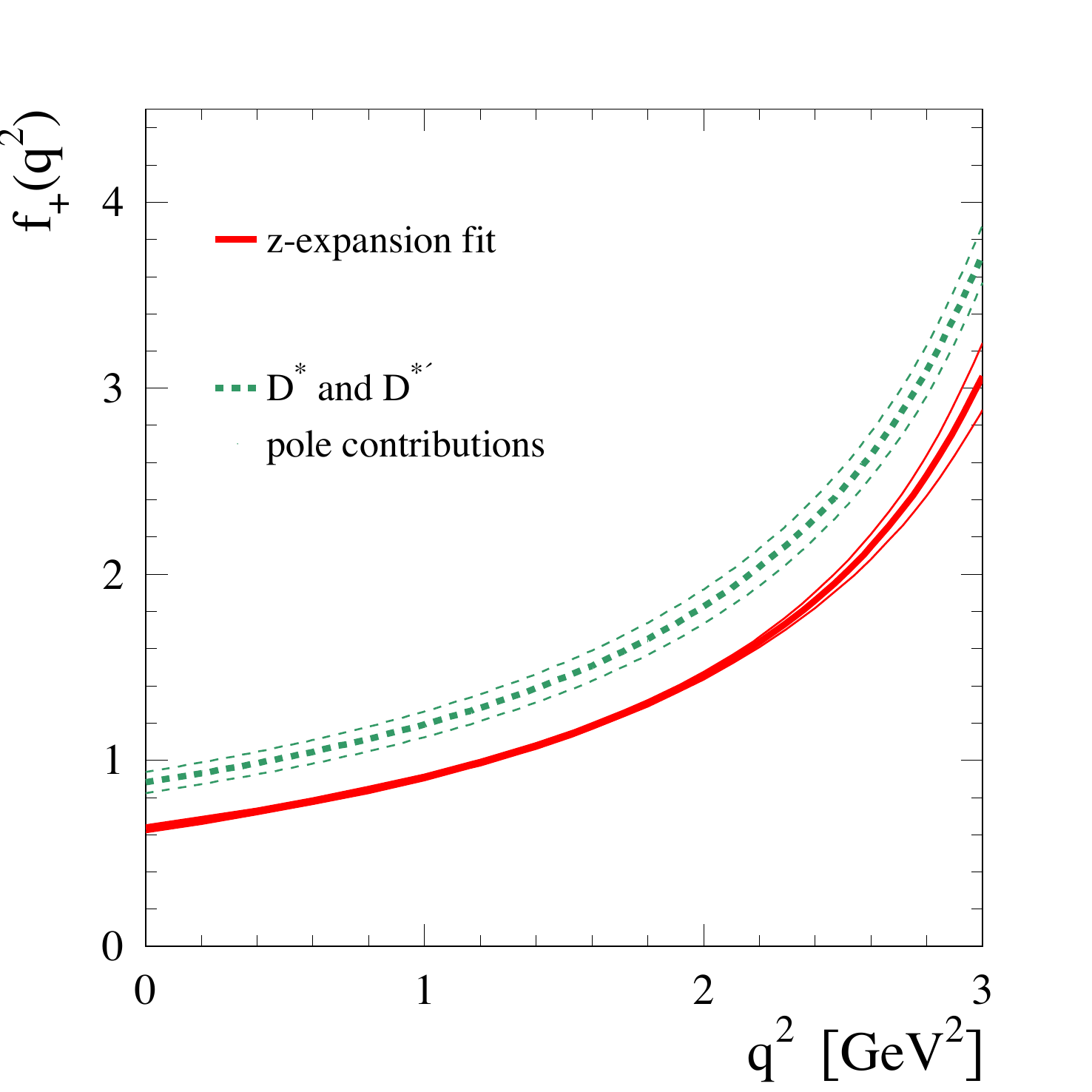}
  \end{center}
  \caption[]{\footnotesize One pole (left plot) and two pole (right plot) shapes of the form factors $f_+^{D\pi}(q^2)$ are compared to the form factor behavior obtained by fitting the experimental data to a so-called $z$-parameterization (cf. Appendix). 
  The one (two) pole form correspond to retaining one (two) term(s) in eq.~(\ref{3poles}), with masses and residua fixed to their physical values, as discussed in the text.  }
\label{fig:patrick0}
\end{figure}

In fig.~\ref{fig:patrick0} we illustrate the behavior of the form factor described by one or two nearest poles, with masses and residua fixed to the values discussed in sec.~\ref{sec-res1} and sec.~\ref{sec-res2} of the present paper, and compare those functions to the form factor extracted from experimental data of 
all four collaborations (CLEO-c, Belle, BaBar, and BESIII, cf. Appendix). 
Clearly the form factor is not described by the first two poles. Since we do not know the properties for the whole series, we can model the form factor by truncating the series beyond the third pole, and promote the third to an effective pole. In other words we model the discontinuity of the form factor beyond the position of the first radially excited vector state by a pole. This amounts to writing
\bea\label{3polesmodel}
f_+(q^2) ={ \gamma_0 \over  m_{D^\ast}^2 - q^2} + { \gamma_1 \over  m_{D^{\ast\prime}}^2 - q^2}+ { \gamma_{\rm eff} \over  m_{\rm eff}^2 - q^2}\,.
\eea
The mass $m_{\rm eff}$ and residue $\gamma_{\rm eff}$ will be determined from the fit with the data for the semileptonic form factor $f^{D\pi}_+(q^2)$. If the form factor was indeed saturated by its third pole, the value of $m_{\rm eff}$ would be the one corresponding to the state with $L=2$ ($m_{D_2^\ast}=m_2=2.82$~GeV), or perhaps to a possible mixture with the higher radial excitation, i.e. state with radial number $n=3$ and $L=0$ ($m_{D^{\ast\prime\prime}}=m_3= 3.1$~GeV), where in the parentheses we quoted the masses obtained in the quark model of ref.~\cite{GI}. Comparing to $m_{\rm eff}=3.1$~GeV seems more reasonable because the coupling to a $L=2$ state is likely to be smaller.~\footnote{That can be seen in the charmonium systems, where the electronic widths have been measured for the first $6$ states, each having $J^P=1^-$. Using $\displaystyle{\Gamma(\psi_n\to e^+e^-) = \frac{4\pi}{3} \frac{4}{9}\alpha_{\rm em}^2 \frac{f_{\psi_n}^2}{m_{\psi_n}^2}}$, and the measured masses and widths, as well as $\alpha_{\rm em}(m_c)=1/134$, one can extract the decay constants to get 
\bea
&&m_{\psi_n}=\{ 3.097, 3.686, 3.773, 4.039(1), 4.191(5), 4.421(4)\}\ \gev ,\nn\\
&& f_{\psi_n}=\{ 407(6), 290(5), 97(4), 182(18), 135(36), 153(38)
\}\ \mev ,
\nn
\eea
in increasing order of $n$. We see that the coupling to the third state (sitting between the first and second radial excitations), with $n=1$ and $L=2$, is indeed much weaker.}

An important theoretical information about higher states --not on particular states, but on their global, infinite series-- comes indeed from eq.~(\ref{eq:unstr}) and the perturbative behavior~(\ref{pertQCD}). The dispersion relation sets an important condition onto the discontinuity which we can see  by expanding around the asymptotically large $Q^2$ (this is in fact the idea of quark-hadron duality). In that case, the leading term behaves as
\bea
{1\over q^2} \int \displaylimits_{t_0}^\infty  dt \ {\rm Im} f_+(t) \,. 
\eea
Since the form factor asymptotically falls faster than $1/q^2$, one obtains that 
\bea
\int \displaylimits_{t_0}^\infty dt\  {\rm Im} f_+(t) = 0 \,,
\eea
which, in our narrow width approximation, results in a condition for  residua, namely, 
\bea\label{superconvergence}
\sum_{i=0}^\infty \gamma_i =0\,.
\eea
This condition is often referred to as {\it superconvergence} and in the case of heavy-to-light decays it was first invoked in ref.~\cite{burdman}. Note however that the integral $\int \displaylimits_{t_0}^\infty dt\  t\  {\rm Im} f_+(t) $ is not convergent and therefore the values $|\gamma_n|$ do not decrease very fast with $n$.

A similar (or equivalent) way to arrive to the superconvergence relation can be made by using heavy quark scaling laws~\cite{Isgur-Wise} in which the heavy-light meson decay constants behave as 
\bea
f_{H^{\ast}_n}= {\hat f_n\over m_h^{1/2}} \,, \qquad g_{H^{\ast}_n H\pi} = m_h g_n \,,
\eea
where $m_h$ stands for the heavy quark mass (either $c$ or $b$), $H^{(\ast)}$ is a corresponding hadronic heavy-light meson, whereas $\hat f_n$ and $g_n$ are constants. 
The above expressions are subjects to the corrections $\propto 1/m_h$ and higher. Therefore, the residua behave as 
\bea \label{scalingRes}
\gamma_n = \hat f_n \ g_n \ m_h^{3/2}\,.
\eea
In the large recoil region for the outgoing pion, which is equivalent to $q^2=0$ in the heavy quark limit, one should verify the scaling law $f_+(0)  \sim m_h^{-3/2}$~\cite{chernyak,leet}, which --when applied to our representation of the form factor as a sum of poles-- translates to, 
\bea
f_+(0) = \sum_i {\gamma_i\over m_i^2} = \sum_i { \hat f_i \ g_i \over m_h^{1/2}} \, \biggl[ 1 + {\cal O}(1/m_h^n) \biggr]  \sim m_h^{-3/2} \,.
\eea 
To ensure the validity of this scaling law, it is therefore necessary that the leading term (proportional to $m_h^{-1/2}$) vanishes, which implies the condition 
\bea
 \sum_{i=0}^\infty  { \hat f_i \ g_i } = 0\,.
\eea 
This last condition is equivalent to the superconvergence relation written above~(\ref{superconvergence}), except that it is taken in the infinite mass limit. In the fits for the effective models, we shall rather use the previous superconvergence at finite mass. 

Obviously, the superconvergence should be treated carefully as the sum might be only slowly converging to zero and many terms might be needed to actually verify eq.~(\ref{superconvergence}). However, since the third pole in our sum~(\ref{3poles}) plays a role of an effective pole, i.e. all the other singularities are represented by an effective pole the position of which ($m_{\rm eff}^2$) can be obtained from the fit with the data, this condition can be legitimate to use as well. To further elucidate this issue we will analyze the experimental data both by using the superconvergence relation~(\ref{superconvergence}) and without it. 

Finally, let us note the following. If we consider together the $B$ and the $D$ decays, we can constrain the fits by using the scaling equation for the residua, eq.~(\ref{scalingRes}), which amounts to assuming that the ratio of the residues for $c$ and $b$ are the same. This is of course true up to $1/m_h$ corrections.This is certainly a strong assumption for the $c$ and even for the $b$ favored heavy-light mesons, which we correct somewhat as explained in the sec.~(\ref{fitting}) in the fits. Moreover, one assumes that this property can also be imposed on the residue of the effective pole.

\subsection{Effect of the hadron loops \label{loops}}
In the above discussion, and in our parameterization~(\ref{3poles}) in particular, we ignored the width of the excited states in denominators, as well as contributions from the cut which have the same origin in loop effects. 
In terms of QCD, ignoring them would mean an implicit assumption of the limit of large number of colors, $N_c\to \infty$, or a flavorless theory, $N_{\rm f}=0$.  However, we know that for higher states 
multi-particle decay channels open up and those states are broad. The effects of hadron loops is manifested by a continuous cut starting from the $D\pi$ threshold, $t_0$. These effects have been tentatively treated by the authors of ref.~\cite{burdman} by inserting approximations to the imaginary part in the above dispersion representation. A more direct approach is to use the analyticity FSI statement (a generalization of the  Watson theorem) which states that the denominator representing singularities of the form factor $D(q^2)$ is the same as the one of the scattering matrix. Namely, we use as a better approximation the analytic generalization of the Breit-Wigner form, with the energy dependent width. What is also crucial is to perform analytic continuation into the physical region. Both are easily performed
through the use of dispersion relations.

Let us consider a given pole as isolated. To be specific, we take the simplest case of a stable particle (for example $m_{B^\ast}^2$ in the case of $B\to \pi\ell\nu_\ell$ decay). The effect is then to add a cut contribution deforming the simple pole form $f_+(q^2)=\gamma_0 /(m_{B^\ast}^2 - q^2)$. Instead, 
\bea
f_+(q^2)={\gamma_0 \over D(q^2)}\,,
\eea
where the denominator $D(q^2)$  can be written as
\bea
D(q^2) = m_{B^\ast}^2 - q^2 - {(m_{B^\ast}^2 - q^2)^2\over \pi} \int \displaylimits_{t_0}^\infty {m_{B^\ast} \Gamma(t)\over (t- m_{B^\ast}^2)^2 (t-q^2)} dt\,.
\eea
The dispersion relation is here subtracted twice to display the difference with $D(q^2) =m_{B^\ast}^2 - q^2$. The result obviously depends on the shape of $\Gamma(t)$ taken into consideration.~\footnote{For instance one can take the form by Gounaris-Sakurai $\Gamma(t) \propto k(t)^3/\sqrt{t}$, where $k(t)=\lambda^{1/2}(t,m_D^2,m_\pi^2)/(2\sqrt{t})$.}  
The naive expression~\cite{gounaris} cannot be valid far away from the pole and therefore one must cut off the integration in some way. Either one can cut the integration interval, like it was done in ref.~\cite{burdman}, or use an attenuation factor \`a la Blatt-Weisskopf~\cite{PDG}.  
We checked that for this nearest pole, and with a reasonably placed cut (a few hundreds MeV away from the pole), the resulting correction to the simple pole-like shape is very small even at $q^2=0$.

\section{Fitting the experimental data~\label{fitting}}

Differential decay rate for $D \rightarrow \pi \ell^+ \nu_{\ell}$ has been accurately measured in  the experiments by Belle~\cite{belleD}, 
CLEO-c~\cite{cleo-c, cleo-c2}, BaBar~\cite{babarD}, and BESIII~\cite{bes3D} . 
From the information provided in their papers we were able to reconstruct the shapes of the form factors and then combine both the Belle, CLEO-c, 
BaBar, and BESIII results to test the three-pole model discussed above. 
Since the published values and the associated uncertainties are provided with limited accuracy, we compare the results we obtain by fitting the published data to the same parameterization used in these papers (see Appendix for more details).
Let us briefly summarize the experimental situation.
\begin{itemize}
\item Belle measured $D^0 \rightarrow \pi^- \ell^+ \nu_{\ell}$ decays with 
$\ell=e$ or $\mu$, using  a $282\,{\rm fb}^{-1}$ data sample. The $D^0$ flavor and momentum are tagged  through a full reconstruction of the recoiling charm meson and additional mesons
from fragmentation. The reconstruction method provides a very good resolution
in neutrino momentum and in $q^2=(p_{\ell}+p_{\nu_{\ell}})^2$. They provide $f_{+}^{D\pi}(q^2)$ with uncertainties at ten values of $q^2$~\cite{belleD}.
\item CLEO-c  measured the decays 
$D^0 \rightarrow \pi^- e^+ \nu_e$ and 
$D^+ \rightarrow \pi^0 e^+ \nu_e$ 
in events tagged by a $D^0$ or a $D^+$ exclusively reconstructed. 
Analyzing an integrated luminosity of $818\,{\rm pb}^{-1}$, they provide the values for partial decay widths in seven $q^2$ bins, for each channel,
together with the corresponding statistical and systematic error matrices~\cite{cleo-c2}.
\item CLEO-c also reported results for the partial widths by analyzing events where the other $D$ meson is not reconstructed~\cite{cleo-c}. 
Although for that analysis they used about a third of their data sample ($281\,{\rm pb}^{-1}$) the results are competitive in quality with those obtained in the tagged analysis despite a higher combinatorial background. The reason for such a good quality is due to 
the reconstruction efficiency of their analysis which is larger by a factor of $2.5$ than in the tagged analysis. 
Values of partial decay branching fractions in five $q^2$ bins are provided 
for each channel with the corresponding statistical and systematic error matrices.

\item 
The BaBar experiment has studied the $q^2$ dependence of the Cabibbo 
suppressed semileptonic decay rates and measured the ratio
$R_D = {\cal B}(D^0 \rightarrow \pi^- e^+ \nu_e)/{\cal B}(D^0 \rightarrow K^- \pi^+)$. 
Taking into account the known branching fraction ${\cal B}(D^0 \rightarrow K^- \pi^+)$, they derive the value of the $D^0 \rightarrow \pi^- e^+ \nu_e$ 
branching fraction.  
The analysis exploits the large production of charm mesons via the process $e^+ e^- \to c\overline{c}$ and identifies $D^0$ from the decay $D^{*+} \to D^0 \pi^+$.
The data amount to a total integrated luminosity of $347.2~{\rm fb}^{-1}$
at a center of mass energy near the $\Upsilon(4S)$ mass. 
Preliminary values of the partial decay branching fraction are measured
in ten $q^2$ intervals~\cite{babarD}.

\item 
Analyzing an integrated luminosity of $2.9\,fb^{-1}$, the BESIII experiment 
has measured the decay
$D^0 \rightarrow \pi^- e^+ \nu_{e}$~\cite{bes3D}
in events tagged by a $\overline{D}^0$ exclusively reconstructed.
Preliminary values of partial decay widths are provided in 
fourteen $q^2$ intervals, along
with corresponding statistical and systematic error matrices.

\end{itemize}
In ref.~\cite{cleo-c3}, CLEO-c combined the tagged and untagged measurements obtained with the same
integrated luminosity of $281\,{\rm pb}^{-1}$ and reported the partial decay rates for several $q^2$-bins for both channels ($D^0 \rightarrow \pi^- e^+ \nu_e$ and 
$D^+ \rightarrow \pi^0 e^+ \nu_e$). In that study they also included the correlation matrices. We transform these matrices to correspond to the larger integrated 
luminosity of $818\,{\rm pb}^{-1}$, analyzed in the tagged 
measurement. Correlation matrices for statistical uncertainties 
on measured branching fractions in the given $q^2$ bins are scaled
by the quantity $\rho_{818,281}$, defined as
\beq\label{eq:LG}
\rho_{818,281} = \rho_{281,281} \frac{\sigma^{\rm tag,stat}_{818}}{\sigma^{\rm tag,stat}_{281}},
\eeq
where $\rho_{281,281}$ is a published correlation for statistical uncertainties 
between tagged and untagged measurements done with the same, $281\,{\rm pb}^{-1}$,
integrated luminosity. $\sigma^{\rm tag,stat}_{818}$ is the statistical uncertainty of the
tagged analysis measurement obtained with $818\,{\rm pb}^{-1}$
integrated luminosity, and $\sigma^{\rm tag,stat}_{281}$ is the statistical uncertainty of the
tagged analysis measurement corresponding to $281\,{\rm pb}^{-1}$. As for systematic uncertainties, we use the  published correlation matrices
between the two analyses~\cite{cleo-c3}.

\subsection{Three pole fit of the $D\to \pi\ell\nu$ data  }

\begin{figure}[!htbp!] 
  \begin{center}
\includegraphics[height=17cm]{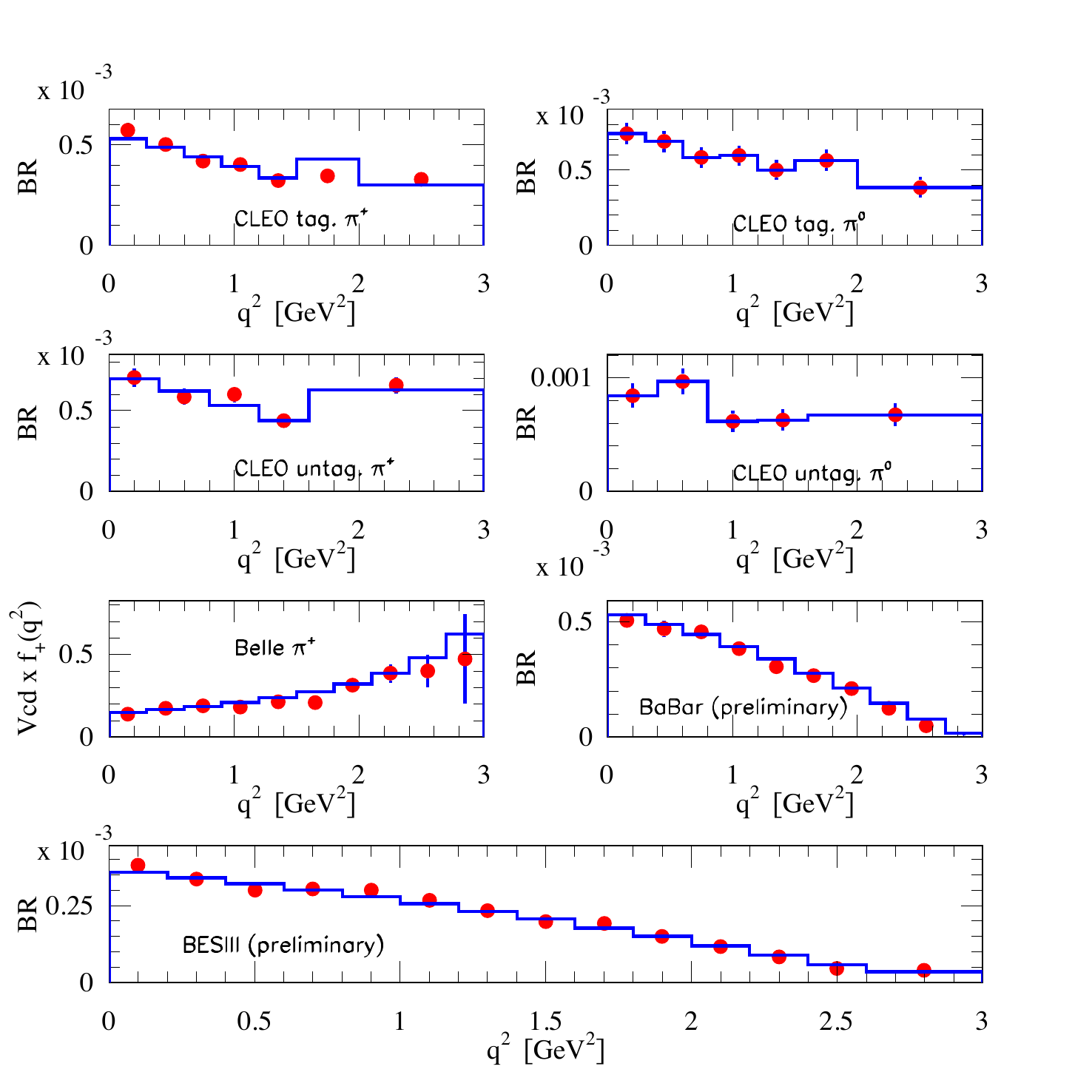}
  \end{center}
  \caption[]{\footnotesize Comparison between $D \rightarrow \pi \ell^+ \nu_{\ell}$ 
measurements (points with error bars)
and the results of the fit of all the data simultaneously (dashed histograms) using the superconvergence
condition and fitting the mass of the third pole. }
\label{fig:patrick1}
\end{figure}

As we showed in fig.~\ref{fig:patrick0} the experimental data cannot be described by the first two physical poles. 
Instead we perform fits to the parameterization~(\ref{3polesmodel})  
in which the residua are either left as free parameters or fixed to the values discussed in the previous sections, namely:
\bea\label{constr}
\gamma_0=4.19(15)~\gev^2,\qquad \gamma_1=-1.0(3)~\gev^2.
\eea
The values of the first two poles are kept fixed $m_{D^\ast}=2.01$~GeV and $m_{D^{\ast\prime}}=2.62$~GeV.  The results of the fits are presented in tab.~\ref{tab:A}. 
\begin{table}[!bp!]
\begin{center}
{
\renewcommand{\arraystretch}{1.7}
{\scalebox{.89}{   \begin{tabular}{|c|ccccc|}
    \hline\hline
{\color{blue} constraint }& {\color{blue}  $\chi^2/{\rm ndf}$} & {\color{blue} $\gamma_0\,[\gev^2]$}&{\color{blue} $\gamma_1\,[\gev^2]$}  &{\color{blue}  $\gamma_{\rm eff}\,[\gev^2]$ } &{\color{blue}  $m_{\rm eff}\,[\gev^2]$ }\\
\hline\hline
$\gamma_{\rm eff}$ free, $m_{\rm eff} = m_{D^{\ast \prime\prime}}$   & $55.7/57$ &$3.95\pm0.09$ & $-1.10\pm 0.29$ & $-1.74\pm0.44$ & $3.1$  \\ \hline
$\gamma_{\rm eff}$ from eq.~(\ref{superconvergence}), $m_{\rm eff} = m_{D^{\ast \prime\prime}}$    & $92.0/58$ &$4.30\pm0.07$ & $-0.19\pm 0.25$ & $-\gamma_0 -\gamma_1$ &$3.1$ \\ \hline
$\gamma_{\rm eff}$ from eq.~(\ref{superconvergence}), $m_{\rm eff}$ free    & $58.0/57$ &$3.88\pm0.09$ & $-1.18\pm 0.30$ & $-\gamma_0 -\gamma_1$ &$4.17\pm0.42$\\
\hline

  \end{tabular}
  }}
  }
 \caption {\footnotesize Results of the fit of the experimental data for the form factor $f_+^{D\pi}(q^2)$ to the 3-poles model in eq.~(\ref{3polesmodel}). We do not present the results of the fits to the one-pole or two-pole formulas, with both masses and residua 
fixed to their physical values. These two situations fail to describe well the experimental data as shown in fig.~\ref{fig:patrick0}.  Eq.~(\ref{superconvergence}) refers to the super convergence conditions discussed in the text. }
\label{tab:A}
\end{center}
\end{table}
We see that the good quality of the fits is obtained if either the mass of the third pole is fixed to its physical value~\cite{GI} while its residuum is left free, or if the superconvergence condition is imposed on residua while the mass of the third 
pole is left as a free parameter. On the contrary, if both the superconvergence condition is imposed and the mass of the third pole is fixed to its physical value,  $\chi^2$ increases by $36$ units and therefore that fit must be discarded.  
Note, however, that by fixing the third pole to $m_{\rm eff}=m_{D^{\ast \prime\prime}} = 3.1$~GeV we obtain the residuum of the third pole larger than the second one. 
In the case  in which the superconvergence condition~(\ref{superconvergence}) is imposed the residuum is also large with respect to the second one, but this time $\gamma_{\rm eff}$ corresponds to the sum of all other contributions 
and it should be viewed as a residuum of an effective pole the
position of which is obtained from the fit. Our result shows that the mass of the third pole preferred by the data [$m_{\rm eff}=4.2(4)$~GeV] is larger than the expected one 
[$\approx 3$~GeV]. In other words the form factor is not saturated by the three physical poles alone, but the effective three-pole picture with the superconvergence condition 
gives us a fully acceptable description of the experimental data if that last pole is treated as effective and left free. The predicted value for the second radial excitation, $m_{D^{\ast\prime\prime}} \approx 3.1$~GeV
 or the one of the first $L=2$ state $m_{D^{**}}=2.82$~GeV, cannot be accommodated to fit the data as the effective pole mass. That statement is valid to about $6\ \sigma$'s and is in contrast to the initial assumption of ref.~\cite{burdman}.
 Fits to three poles are shown in fig.~\ref{fig:patrick1}. Plots obtained by either imposing the superconvergence condition 
or not are practically indistinguishable. 

\subsection{Three pole fit of the $B\to \pi\ell\nu$ data}

Experimental data for $B^0\to \pi^-\ell^+\nu_\ell$  that we use here are the measurement made at $B$-factories. More specifically,
\begin{itemize} 
\item BaBar used $462 \times 10^6$ $B\overline{B}$ pair events in which 
the signal B decays are reconstructed with 
a loose neutrino reconstruction
technique~\cite{ref:babarb}. By using the isospin symmetry, they combine measurements 
of $B^0 \rightarrow \pi^- \ell^+ \nu_{\ell}$ and  
$B^+ \rightarrow \pi^0 \ell^+ \nu_{\ell}$, and express the partial branching fractions in twelve bins in terms of the decay
$B^0 \rightarrow \pi^- \ell^+ \nu_{\ell}$ ($\ell= e$ or $\mu$).  
\item Belle used a data sample
containing $657\times 10^6$  $B\overline{B}$ events and present results in thirteen bins for the  decay
$B^0 \rightarrow \pi^- \ell^+ \nu_{\ell}$~\cite{ref:belleb}.
\end{itemize}
We combine the data obtained by the two experiments and fit them simultaneously to a form analogous to eq.~(\ref{3poles}),  
\begin{align}\label{3polesB}
f_+^{B\pi}(q^2) &={ \beta_0 \over  m_{B^\ast}^2 - q^2} + { \beta_1 \over  m_{B^{\ast\prime}}^2 - q^2}+ { \beta_{\rm eff} \over  m_{\rm eff}^2 - q^2}  \nn\\
 &={ \beta_0 \over  m_{B^\ast}^2 } \left(  {1\over 1- q^2/m_{B^\ast}^2} +  { \beta_1 \over  \beta_0} {m_{B^\ast}^2 \over m_{B^{\ast\prime}}^2} {1\over 1- q^2/m_{B^{\ast\prime}}^2}
 +
  { \beta_{\rm eff} \over  \beta_0} {m_{B^\ast}^2 \over m_{\rm eff}^2} {1\over 1- q^2/m_{\rm eff}^2} 
\right) \nn\\
&={ \beta_0 \over  m_{B^\ast}^2 } \left(  {1\over 1- q^2/m_{B^\ast}^2} +     {R_1\over 1- q^2/m_{B^{\ast\prime}}^2}
 + 
  {R_2 \over 1- q^2/m_{\rm eff}^2} 
\right) \,,
\end{align}
where $\beta_0$ and $\beta_1$ are the residua discussed in the previous sections [cf. eqs.~(\ref{eq:resB},\ref{eq:res5})], namely
\bea\label{eq:resBB}
\beta_0=(24.3\pm 3.4)~\gev^2,\quad \beta_1=-(8\pm 2)~\gev^2,
\eea
which together with the masses $m_{B^\ast}=5.325$~GeV and $m_{B^{\ast\prime}}=5.97(1)$~GeV,  gives  
\bea\label{eq:resBBbis}
{ \beta_0 \over  m_{B^\ast}^2 } =0.86\pm 0.12\,,\quad R_1=- (0.25\pm 0.06)\,.
\eea
We can now perform a fit of the experimental data for $\vert V_{ub}\vert f_+(q^2)$, extracted from the partial branching fractions, to eq.~(\ref{3polesB}) by using the above values as constraints, and vary the value of $m_{\rm eff} > m_{B^{\ast \prime}}$ to extract $R_2$, which is left as a free parameter.~\footnote{Note that the expression for the differential decay rate for $B\to \pi\ell\nu_\ell$ is of the same form as eq.~(\ref{eq:ddr}), in which one replaces $V_{cd}\to V_{ub}$ and $m_D\to m_B$.} In doing so, we also let  $\vert V_{ub}\vert \beta_0/m_{B^\ast}^2$ as a free parameter and only in the end we can divide by $\beta_0/m_{B^\ast}^2$, given in eq.~(\ref{eq:resBBbis}), to practically get a value for $\vert V_{ub}\vert$. The results of such fits are shown in tab.~\ref{tab:pp2}.
\begin{table}
\begin{center}
\hspace*{-4mm}{
\renewcommand{\arraystretch}{1.7}
  \begin{tabular}{|cccccc|}
    \hline\hline
$m_2\,[\gev]$& $\chi^2$ & $|V_{ub}|\times \beta_0/m_{B^\ast}^2\times 10^2$&$R_1$& $R_2$ & $1+\beta_1/\beta_0+\beta_2/\beta_0$\\
\hline
$6$ & $31.0$ & $7.91\pm0.33$ & $-0.26\pm0.06$ & $-0.40\pm0.07$ & $+0.17\pm 0.02$\\
$7$ & $24.8$ & $6.83\pm0.29$ & $-0.24\pm0.06$ & $-0.37\pm0.06$ & $+0.06\pm 0.04$\\
$8$ & $22.5$ & $6.50\pm0.29$ & $-0.24\pm0.06$ & $-0.36\pm0.06$ & $-0.11\pm 0.06$\\
$10$ & $20.7$ & $6.26\pm0.29$ & $-0.24\pm0.06$ & $-0.35\pm0.05$ & $-0.52\pm 0.12$\\
$12$ & $20.0$ & $6.17\pm0.29$ & $-0.25\pm0.06$ & $-0.34\pm0.05$ & $-1.0\pm 0.2$\\
$15$ & $19.6$ & $6.10\pm0.30$ & $-0.25\pm0.06$ & $-0.34\pm0.05$ & $-2.0\pm 0.3$\\
$30$ & $19.2$ & $6.03\pm0.30$ & $-0.25\pm0.06$ & $-0.33\pm0.05$ & $-9.8\pm 1.5$\\
\hline\hline
  \end{tabular}
}
    \caption {\footnotesize Results obtained with a 3-poles model for 
$f_{+}^{B\pi}(q^2)$ and for different values of the third pole mass.
The total number of degrees of freedom is equal to 23.
The superconvergence condition is not used. 
Untagged analyses by BaBar and Belle are used.}
\end{center}
\label{tab:pp2}
\end{table}
In the same table we also compute $1+\beta_1/\beta_0+\beta_{\rm eff}/\beta_0$ which should be equal to zero in the case the superconvergence relation~(\ref{superconvergence}) is verified by the fit. 
We see that this indeed happens for $m_{\rm eff}\in (7,8)$~GeV. That last feature happens also in the case in which we choose $R_1$ to be the value from the $D\to \pi \ell \nu_\ell$, 
namely $R_1= R_1^D=-0.15(5)$. From these fits we get $|V_{ub}|\simeq 3\times 10^{-3}$, and the value is smaller for smaller values of $R_1$. 

The choice $R_1^B= R_1^D$ was actually first proposed in ref.~\cite{burdman} where it was argued that the ratio of residua for two different poles is independent on the heavy quark mass, as a consequence of the scaling behaviour~(\ref{scalingRes}). 
By using eq.~(\ref{DtoB}) and a complete set of decay constants given in ref.~\cite{Gelhausen:2014jea} one indeed verifies that $R_1^D=-0.19(5)$ is fully compatible with that of the $B$-mesons. In our results for $R_1^D=-0.15(5)$ and $R_1^B=-0.25(6)$ the central values are somewhat different but the results are obviously compatible.~\footnote{The difference in the central values comes from the fact that for our estimate of $R_1^D$ we use the SU(6) motivated relation, in connection with the lattice QCD result for  $f_{D^\prime}/f_{D}$, to get $f_{D^{\ast\prime}}/f_{D^\ast} = (f_{D^\prime}/f_{D})^{\rm latt} (f_{D^\ast}/f_{D})^{\rm latt} \times f_D^{\rm exp}$, which is compatible with the QCD sum rule value, even if the central value is smaller [cf. discussion after eq.~(\ref{ref:s4})].}

Equivalently, we can also impose the validity of the superconvergence~(\ref{superconvergence}) and again perform various fits using (parts of) the physical information from eq.~(\ref{eq:resBBbis}). 
\begin{table}[!t!]\label{tab:p1}
\begin{center}
{
\renewcommand{\arraystretch}{1.7}
\hspace*{-9mm}  \begin{tabular}{|c|ccccc|}
    \hline\hline
{\color{blue} constraint }& {\color{blue}  $\chi^2/{\rm ndf}$} & {\color{blue} $\beta_0\,[\gev^2]$}&{\color{blue} $\beta_1\,[\gev^2]$}  &{\color{blue} $m_{\rm eff}\,[\gev]$ } &{\color{blue}  $|V_{ub}|\times 10^3$ }\\
\hline\hline
$\beta_0$ eq.~(\ref{eq:resBB}), $\beta_1$ free& $22.9/22$ &$24.3\pm 3.4$ & $\simeq 0$ & $7.0\pm0.2$ &$2.56\pm0.39$  \\ 
\hline
$\beta_0$ and $\beta_1$ eq.~(\ref{eq:resBB})  &$23.7/23$ & $24.6\pm3.4$ & $-7.7\pm2.0$ & $7.4\pm0.3$ & $2.70\pm0.42$  \\
\hline
$\beta_0$ eq.~(\ref{eq:resBB}),$\beta_1=-(16\pm4)\ \gev^2$   &$25.3/23$ &$25.3\pm 3.3$ & $-13.5\pm 3.4$ & $8.2\pm0.8$ & $2.79\pm0.43$  \\ \hline
\hline

  \end{tabular}
  }
 \caption {\footnotesize {Results obtained with a 3-poles model for 
$f_+^{B\pi}(q^2)$ and for different constraints on the residues at the
$B^*$ and $B^{*\prime}$ poles. The superconvergence condition is used ($\beta_2\equiv \beta_{\rm eff}=-\beta_0-\beta_1$).
Measurements from untagged BaBar and Belle samples are analyzed.}}
\end{center}
\label{tab:p2}
\end{table}
The corresponding results are shown in tab.~\ref{tab:p2} where we also quote the values for $\vert V_{ub}\vert$ obtained from these fits, which remain slightly below but compatible with $0.003$ typically extracted from the exclusive decay modes, and certainly lower than the value extracted from the inclusive $B\to X_u\ell\nu_\ell$ decays. It is important to emphasize, however, that the main part of uncertainty comes from theory. We get
\bea
\vert V_{ub}  \vert = (2.7\pm 0.1_{\rm exp}\pm 0.4_{\rm theo})\times 10^{-3},
\eea
where most of the theory error comes from $g_b$ and $f_{B^\ast}$ while the impact of uncertainty in the residuum $\beta_1$ is small compared to that of $\beta_0$.
\begin{figure}[!htb]
  \begin{center}
\hspace*{-1cm}\includegraphics[width=.55\textwidth]{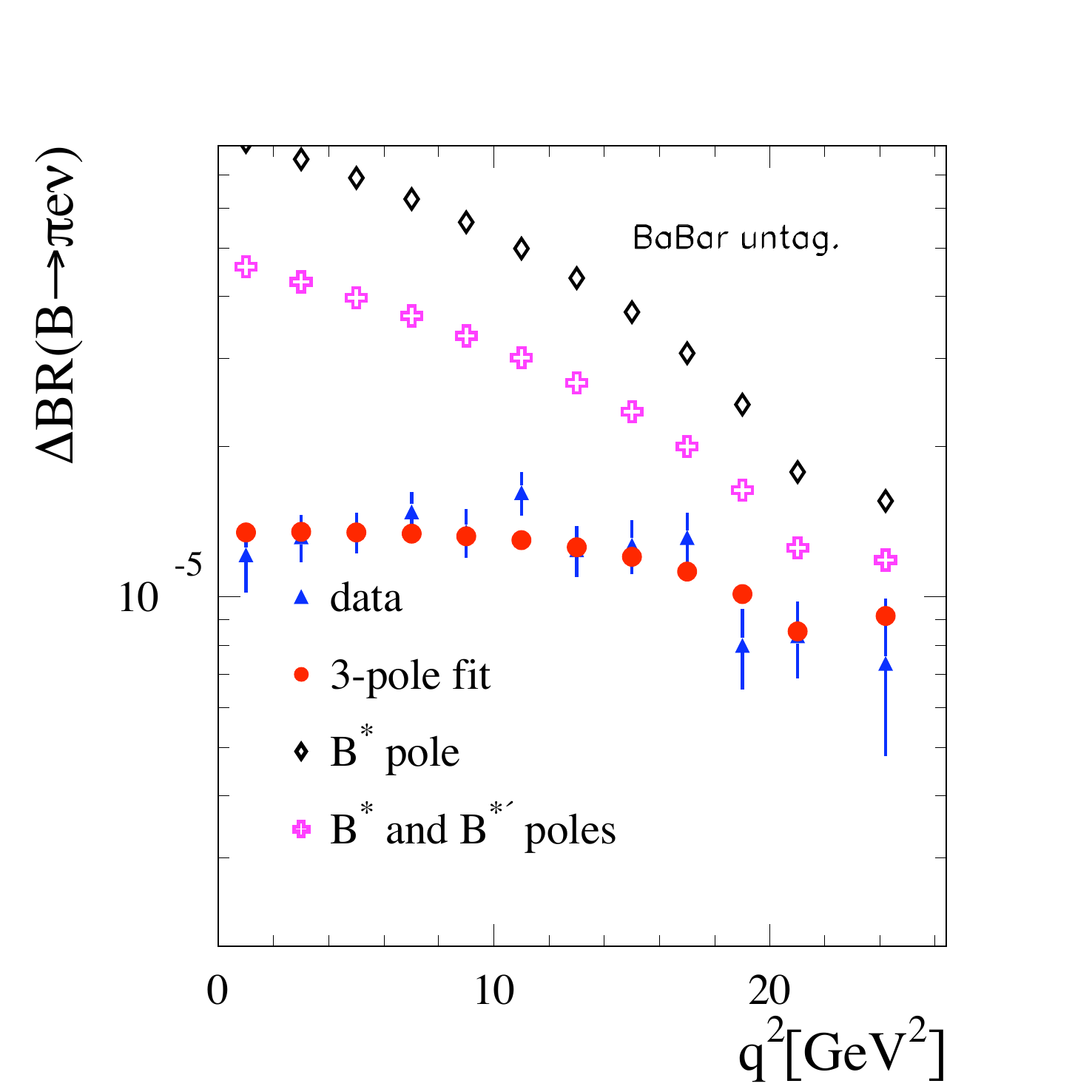}~\includegraphics[width=.55\textwidth]{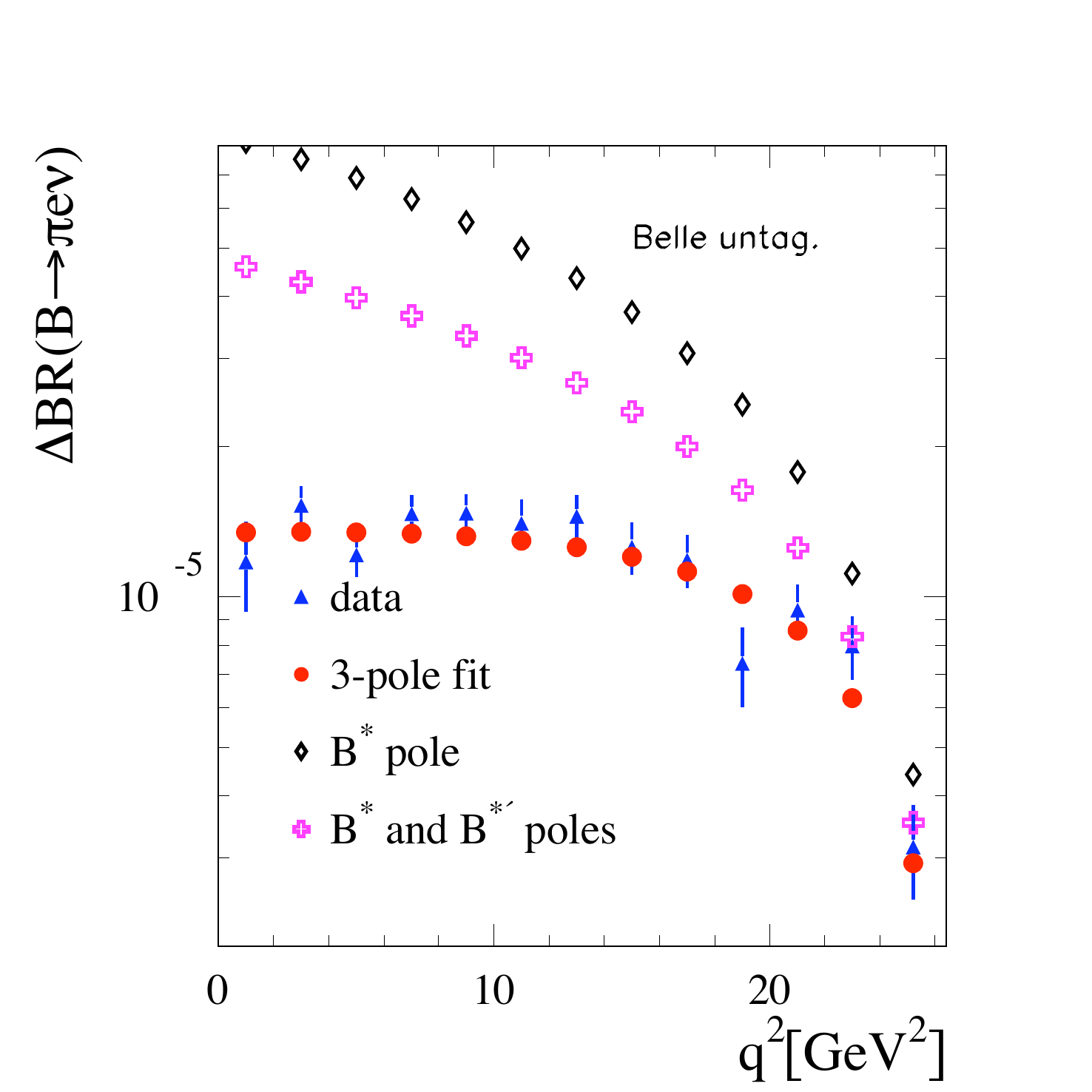}
  \end{center}
  \caption[]{ \footnotesize Comparison between the measured partial
decay rate in several $q^2$ bins for $\Bdecpi$ in BaBar and Belle
untagged analyses, 
with expectations using a 3-poles model. Contributions from
the $B^*$ pole and from the sum of the $B^*$ and $B^{*\prime}$ poles
are displayed also.}
  \label{fig:dbdq2_3poles}
\end{figure}

\subsection{Summary and Conclusion\label{sec:concl}}

In this paper we tested the idea of describing the vector semileptonic form factor $f_+(q^2)$ in terms of a series of poles. Using the recent experimental information on the $q^2$ spectrum of $D\to \pi \ell\nu_\ell$ decay on one hand, and the information on the residuum of the form factor at its first two poles, we could extend the parameterization proposed in ref.~\cite{BK} and test a model with three poles in which the third pole is only effective, i.e. it parameterizes the singularities beyond the second pole. From the fit with data we found that the position of that effective pole is around $m_{\rm eff}\approx 4$~GeV, which is larger than the value of the second radial excitation of the $D^\ast$ meson which is expected to have a mass of about $3.1$~GeV. In other words we actually exclude the possibility that the form factor can be saturated by the three first physical poles alone. In arriving to that conclusion we checked and then relied on the superconvergence condition which basically means that the sum of all residua in the pole-models is equal to zero. That rule comes from the dispersion relation for the form factor, combined with the asymptotic QCD behavior of the form factor, and can also be obtained on the basis of the heavy quark expansion considerations. 

It should be emphasized that the physical relevance of our effective pole model, which could be questioned because of the empirical character of the effective pole, is strongly supported by the fact that one checks the expected superconvergence and the heavy quark scaling properties, as explained in sec.~\ref{super}, even without having to assume them a priori. 

We then applied the same considerations to the $B\to \pi \ell\nu_\ell$ which is particularly interesting for the extraction of $\vert V_{ub}\vert$. We found that the precision of the extracted $\vert V_{ub}\vert$ is mostly plagued by the residuum of the form factor at its first pole. In other words, a better accuracy of the lattice QCD determination of $g_{B^\ast B\pi}$ (i.e. $g_b$) could be very beneficiary for a more precise  extraction of $\vert V_{ub}\vert$ from $B\to \pi \ell\nu_\ell$.

For the case of $B$ decay we observe that the third (effective) pole is situated at $m_{\rm eff}\approx 7$~GeV, when the  superconvergence condition is imposed. This is approximately $2$~GeV above the lowest lying vector meson $B^\ast$, which is practically 
equal to the similar mass difference in the case of $D$ decay. It can be shown that this equality is just what is required if we impose the $m_h^{-3/2}$  scaling to the effective pole model~\cite{chernyak,leet}. 
The size of that difference, however, cannot be predicted by our simple treatment. The relatively large value we find ($\simeq 2$~GeV) indicates that many states are contributing to the effective pole. That latter statement is also supported by the size of the effective residuum which should be smaller if it were saturated only by a third (or/and fourth) pole.

In fixing the residuum of both the $D\to\pi$ and $B\to\pi$ form factors at their first respective pole, we needed an accurate estimate of the vector meson decay constant. We provided the values for $f_{D^\ast}/f_D$ and $f_{B^\ast}/f_B$ by using the numerical simulations of QCD on the lattice. The former value is an improvement with respect to the previously published number, while the latter is the first unquenched lattice QCD estimate.  

In testing the idea of parameterizing the form factor by three poles, we used 
the currently available published data by CLEO-c, Belle, BaBar, and BESIII for 
$D\to \pi\ell\nu_\ell$, and those obtained by BaBar and Belle for $B\to \pi\ell\nu_\ell$ decay.  These data can be used to extract the information on the $D\pi$ ($B\pi$) scattering following the idea of ref.~\cite{nieves}.

\section*{Acknowledgments}
We thank L.~Gibbons for discussions concerning the treatment of experimental data, G.~Rong for allowing us to use the preliminary BESSIII results in our analysis, the members of the ETMC, and GENCI (2013-056808) for according us computing time at IDRIS Orsay to perform the lattice QCD calculations needed for this work.

\newpage
\appendix
\renewcommand\thetable{A.\arabic{table}}
\setcounter{table}{0}
\section{Results for the ratio of the vector to pseudoscalar meson decay constants directly accessed on the lattices \label{app1}}

In this appendix we present the result for the ratio of the decay constants $f_{H^\ast}/f_H$ as obtained on the lattices used in this work (cf. tab.~\ref{tab:0}) and for a series of seven heavy quark masses, starting from the charm quark and then successively increased by a factor of $\lambda = 1.175$. The results for $f_H$ and $f_{H^\ast}$ are obtained from the fit of correlation functions to the forms indicated in eq.~(\ref{r1}) and by using the definitions given in eq.~(\ref{r2}).

\begin{table}[h!]
\begin{center}
\hspace*{-7mm}{
\renewcommand{\arraystretch}{1.8}
{\scalebox{.89}{  \begin{tabular}{|c|ccccccc|}
    \hline\hline
$\beta,L, \mu_{\rm sea}$ & $m_c$ & $\lambda m_c$ & $\lambda^2 m_c$ & $\lambda^3 m_c$ & $\lambda^4 m_c$ & $\lambda^5 m_c$ & $\lambda^6 m_c$  \\
\hline
$3.80,24,0.0080$ & $1.221(23)$ & $1.227(29)$ & $1.235(29)$ & $1.252(32)$ & $1.278(39)$ & $1.313(53)$ & $1.360(80)$\\
$3.80,24,0.0110$ & $1.209(17)$ & $1.210(18)$ & $1.216(20)$ & $1.231(23)$ & $1.256(25)$ & $1.293(27)$ & $1.343(34)$\\ \hline
$3.90,24,0.0040$ & $1.222(18)$ & $1.215(17)$ & $1.214(16)$ & $1.217(15)$ & $1.227(15)$ & $1.245(16)$ & $1.273(21)$\\
$3.90,24,0.0064$ & $1.207(24)$ & $1.205(25)$ & $1.208(26)$ & $1.219(27)$ & $1.242(28)$ & $1.278(29)$ & $1.324(32)$\\
$3.90,24,0.0085$ & $1.192(20)$ & $1.191(21)$ & $1.194(21)$ & $1.202(21)$ & $1.216(22)$ & $1.239(23)$ & $1.281(26)$\\
$3.90,24,0.0100$ & $1.216(14)$ & $1.207(14)$ & $1.202(14)$ & $1.201(15)$ & $1.207(17)$ & $1.223(19)$ & $1.251(21)$\\
\hline
$3.90,32,0.0030$ & $1.177(21)$ & $1.170(21)$ & $1.170(22)$ & $1.203(22)$ & $1.199(28)$ & $1.241(46)$ & $1.284(46)$\\
$3.90,32,0.0040$ & $1.227(16)$ & $1.221(16)$ & $1.222(16)$ & $1.209(14)$ & $1.210(16)$ & $1.215(23)$ & $1.215(35)$\\
\hline
$4.05,32,0.0030$ & $1.204(9)$ & $1.192(9)$ & $1.183(9)$ & $1.179(10)$ & $1.180(10)$ & $1.188(13)$ & $1.207(17)$\\
$4.05,32,0.0060$ & $1.171(23)$ & $1.168(22)$ & $1.167(21)$ & $1.170(20)$ & $1.177(20)$ & $1.190(21)$ & $1.211(22)$\\
$4.05,32,0.0080$ & $1.240(18)$ & $1.225(17)$ & $1.215(17)$ & $1.210(16)$ & $1.213(17)$ & $1.224(19)$ & $1.248(22)$\\
\hline
$4.20,32,0.0065$ & $1.177(32)$ & $1.162(31)$ & $1.147(32)$ & $1.139(35)$ & $1.136(40)$ & $1.139(48)$ & $1.271(143)$\\
\hline
$4.20,48,0.0020$ & $1.268(101)$ & $1.269(100)$ & $1.275(101)$ & $1.287(101)$ & $1.308(104)$ & $1.339(108)$ & $1.381(114)$\\
\hline\hline
  \end{tabular}
}
}}
    \caption {\footnotesize Results for $f_{H^\ast}/f_H$ obtained for all of our heavy quark masses and for all of our lattice ensembles, with the light valence quark mass equal to that of the sea quark. Rows correspond to our $13$ ensembles of the lattice data, while columns correspond to the heavy quark masses we consider in this work ($\lambda=1.175$). }
\end{center}
\end{table}

\renewcommand\thetable{B.\arabic{table}}

\section{Validation of the fitting procedure \label{sec:appendixa}}

Since the experimental measurements and their corresponding
uncertainty matrices are published with limited accuracy, we verify
in the following that the values for model parameters used to fit the data of each experiment agree with the values we obtain by fitting 
the same model on their published measurements.

These comparisons indicate that published and refitted values of model
parameters agree within a fraction of the total uncertainty and that
uncertainties on fitted quantities agree as well.

Apart from the Belle experiment, other collaborations have fitted their data by using the second order $z$-expansion parameterization 
of the hadronic 
form factor. This parameterization is used also in the present analysis 
to obtain an average
of all measurements which was then compared in sec.~\ref{super} with expectations from the $D^*$ pole alone
and from the sum of the $D^*$ and $D^{*\prime}$ poles.

\subsection{$z$-expansion parameterization of  $f_{+}^{D\pi}(q^2)$}

The $z$-expansion of the form factor is based on general properties of analyticity, unitarity and crossing symmetry. Except for physical poles and multiparticle thresholds, the form factors are analytic function of $q^2$ that can be expressed 
as a convergent power series, given a change of variables~\cite{ref:beforehill} of the following form,
\beq
z(t,t_0) = \frac{\sqrt{t_+-t} - \sqrt{t_+-t_0}}{\sqrt{t_+-t} + \sqrt{t_+-t_0}}
\eeq 
where $t_+=(m_{D}+m_{\pi})^2$ and $t_0=t_+(1 - \sqrt{1-t_-/t_+})$
with $t_-=q^2_{\rm max}=(m_{D}-m_{\pi})^2$.
This transformation maps the kinematic region for the semileptonic decay ($0<q^2<t_-$)
onto a segment on the real axis between $\pm z_{\rm max}$ ($=\pm 0.167$).
In terms of the variable $z$, the form factor, consistent with weak constraints from QCD, can be expanded as
\beq
f_{+}^{D\pi}(t)=\frac{1}{P(t) \Phi(t,t_0)}\sum_{k=0}^{\infty}a_k(t_0)~ z^k(t,t_0).
\label{eq:taylor}
\eeq
The expressions for the functions $P(t)$ and $\Phi(t,t_0)$ can be found in the mentioned references, and they are chosen in such a way that 
\beq
\sum_{k=0}^{\infty}a_k^2(t_0)\leq 1 \,.\label{eq:unitary}
\eeq
The first terms in this expansion appear to be numerically small. The parameterization~(\ref{eq:taylor}) remains valid and
it has been compared with measurements, indicating that the first two terms in the expansion are sufficient to describe the actual data~\cite{ref:hill1}.
To improve the stability of the fits, it is convenient to normalize the $z$-expansion coefficients to $a_0$, i.e. to redefine $r_k = a_k / a_0$, $(k = 1,2)$. 
While being well suited to fit the data, the $z$-expansion has also some disadvantages when compared 
to more phenomenological parameterizations~\cite{DescotesGenon:2008hh}. 
Specifically, there is no direct physical interpretation of the coefficients $a_k(t_0)$.
The contribution from the first pole ($D^{*+}$) is difficult to obtain because it requires extrapolation beyond the physical region,
to the value of $z(m_{D^*}^2,t_0)$ and because, apart from the first two, the other coefficients are not constrained by the available data. Such an extrapolation introduces sizable uncertainties.

\subsection{$D^{0} \rightarrow \pi^- \ell^+ \nu_\ell$ measurement in Belle  \cite{belleD}}

Ten values of $f_{+}^{D\pi}(q^2)$ and corresponding total uncertainties
are published in \cite{belleD}. These values were obtained 
using $|V_{cd}|=0.224(12)$
and a $D^0$ lifetime of $(410.3 \pm 1.5)\times 10^{-15}$s.

We use the values for the products 
$f_{+}^{D\pi}(q^2) \times |V_{cd}|$ to compare with model 
estimates and the common uncertainty to each measurement
originating from the uncertainty on $|V_{cd}|$ is removed from the total
uncertainty of each measurement. 
It is assumed that measurements at different
$q^2$ values are uncorrelated. This hypothesis is justified
considering the high experimental $q^2$ resolution.

Parameters of the modified pole model of ref.~\cite{BK}, fitted on the ten measurements
of the hadronic form factor, are compared in tab.~\ref{tab:verif_belle} with the corresponding results published 
by Belle.

\begin{table}
\begin{center}
  \begin{tabular}{lcc}
    \hline\hline
parameters & published values \cite{belleD} & refitted values\\
\hline
$f_{+}^{D\pi}(0) \times |V_{cd}|$ & $0.140(4)(7)$ & $0.139(7)(8)$ \\
$\alpha_P$ & $0.10\pm0.21\pm0.10$ & $0.11\pm0.19\pm0.13$ \\
$\chi^2/{\rm ndf}$ &$6.4/10$ &$5.5/8$\\
\hline\hline
  \end{tabular}
  \caption {\footnotesize Comparison between published and refitted values of the
parameters entering in the modified pole parameterization of the hadronic 
form factor $f_{+}^{D\pi}(q^2)$ obtained with Belle data~\cite{belleD}.}
\end{center}
\label{tab:verif_belle}
\end{table}

\subsection{CLEO tagged analysis  \cite{cleo-c2}}
The CLEO-c collaboration has published values for the partial decay width
of the $D^0 \rightarrow \pi^- e^+ \nu_e$ and
$D^+ \rightarrow \pi^0 e^+ \nu_e$ decay channels in seven $q^2$
intervals and the corresponding $14\times 14$ uncertainty matrices.
Using these values, and taking the $D^0$ and $D^+$ lifetimes from 
\cite{PDG}, in addition to  assuming the isospin symmetry, we performed 
a simultaneous fit of these measurements.
Values of the parameters of the fit to the $z$-expansion 
parameterization are compared 
with the published results in tab.~\ref{tab:verif_cleot}.
The fit quality is given by the ratio $\chi^2/{\rm ndf}=11.9/11$.

\begin{table}
\begin{center}
  \begin{tabular}{lcc}
    \hline\hline
parameters & published values \cite{cleo-c2} & refitted values\\
\hline
$f_{+}^{D\pi}(0) \times |V_{cd}|$ & $0.150(4)(1)$ & $0.150\pm0.004$ \\
$r_1$ & $-2.35(43)(7)$ & $-2.49\pm0.44$ \\
$r_2$ & $3(3)(0)$ & $4.0\pm 2.8$ \\
$\rho_{01},\rho_{02},\rho_{12}$ &$-0.43,\,0.67,\,-0.94$&$-0.43,\,0.67,\,-0.94$\\$\chi^2/{\rm ndf}$ &$10.4/11$ &$11.9/11$\\
\hline\hline
  \end{tabular}
  \caption {\footnotesize Comparison between published and refitted values of the
parameters entering in the $z$-expansion parameterization of the hadronic 
form factor $f_{+}^{D\pi}(q^2)$. Measurements of $D^0$ and $D^+$ channels
obtained in the CLEO-c tagged analysis \cite{cleo-c2} are combined. $\rho_{ij}$ are the correlation coefficients between the fitted parameters.}
\end{center}
\label{tab:verif_cleot}
\end{table}

\subsection{CLEO untagged analysis  \cite{cleo-c}}
In the CLEO untagged analysis ~\cite{cleo-c}, they
obtain the values of the partial decay width, for the $D^0$ and $D^+$ decay
channels, in five $q^2$-intervals.
Values of parameters in the $z$-expansion 
parameterization of the hadronic form factor extracted from the fit are compared 
with published results~\cite{cleo-c} in tab.~\ref{tab:verif_cleout}, separately
for the $D^0 \rightarrow \pi^- \ell^+ \nu_{\ell}$ and 
$D^+ \rightarrow \pi^0 \ell^+ \nu_{\ell}$ decay channels.

\begin{table}
\begin{center}
  \begin{tabular}{lcc}
    \hline\hline
parameters & published values & refitted values\\
\hline
$D^0 \rightarrow \pi^- e^+ \nu_e$& & \\
\hline
$f_{+}^{D\pi}(0) \times |V_{cd}|$ & $0.140(7)(3)$ & $0.1395\pm0.0076$ \\
$r_1$ & $-2.1(7)(3)$ & $-2.08\pm0.75$ \\
$r_2$ & $-1.2(4.8)(1.7)$ & $-1.1\pm 4.9$ \\
$\rho_{01},\rho_{02},\rho_{12}$ &$?,\,?,\,-0.96$&$-0.51,\,0.71,\,-0.96$\\$\chi^2/{\rm ndf}$ &$2.0/2$ &$1.9/2$\\
\hline
$D^+ \rightarrow \pi^0 e^+ \nu_e$& & \\
\hline
$f_{+}^{D\pi}(0) \times |V_{cd}|$ & $0.139(11)(4)$ & $0.1396\pm0.0106$ \\
$r_1$ & $-0.2(1.5)(0.4)$ & $-0.51\pm1.47$ \\
$r_2$ & $-9.8(9.1)(2.1)$ & $-8.7\pm 8.7$ \\
$\rho_{01},\rho_{02},\rho_{12}$ &$?,\,?,\,-0.96$&$-0.47,\,0.66,\,-0.96$\\$\chi^2/{\rm ndf}$ &$2.8/2$ &$2.8/2$\\

\hline\hline
  \end{tabular}
  \caption{\footnotesize Comparison between published and refitted values of the
parameters entering the $z$-expansion parameterization of the hadronic 
form factor $f_{+}^{D\pi}(q^2)$. Measurements of $D^0$ and $D^+$ channels
obtained in the CLEO-c untagged analysis~\cite{cleo-c} are considered separately.}
\end{center}
\label{tab:verif_cleout}
\end{table}

\subsection{$D^{0} \rightarrow \pi^- e^+ \nu_e$ measurement in Babar  \cite{babarD}}

BaBar measured the partial branching fraction of $D^{0} \rightarrow \pi^- e^+ \nu_e$
in ten $q^2$ intervals and provide the corresponding 
statistical and systematic uncertainty matrices.
Using these (preliminary) results we verified that the values of the $z$-expansion
parameters, also given in that publication, can be correctly reproduced.
(see tab.~\ref{tab:verif_babar}).

\begin{table}

\begin{center}
  \begin{tabular}{lcc}
    \hline\hline
parameters & preliminary values & refitted values\\
\hline
$f_{+,D}^{\pi}(0) \times V_{cd}$ & $0.137\pm0.005$ & $0.138\pm0.004$ \\
$r_1$ & $-1.31\pm0.82$ & $-1.42\pm0.80$ \\
$r_2$ & $-4.2\pm4.4$ & $-3.5\pm 4.2$ \\
$\rho_{01},\rho_{02},\rho_{12}$ &$-0.40,\,0.57,\,-0.97$&$-0.40,\,0.57,\,-0.97$\\$\chi^2/NDF$ &$2.0/7$ &$2.2/7$\\
\hline\hline
  \end{tabular}
  \caption {\footnotesize Comparison between to be published and refitted values of the
parameters entering in the $z$-expansion parameterization of the hadronic 
form factor $f_{+,D}^{\pi}(q^2)$ using BaBar measurements of 
$D^{0} \rightarrow \pi^- e^+ \nu_e$.}\end{center}
\label{tab:verif_babar}
\end{table}

\subsection{$D^{0} \rightarrow \pi^- e^+ \nu_e$ measurement in BESIII  \cite{bes3D}}

The BESIII collaboration has obtained preliminary values for the partial 
decay width
of the $D^0 \rightarrow \pi^- e^+ \nu_e$ decay channel, in fourteen
 $q^2$ intervals and corresponding $14\times 14$ uncertainty matrices.
Using these (preliminary) results we verified that the values of the $z$-expansion
parameters, also given in that publication, can be correctly reproduced.
(see tab.~\ref{tab:verif_bes3}).
The quality of this fit is given by the ratio $\chi^2/NDF=12.0/11$.

\begin{table}

\begin{center}
  \begin{tabular}{lcc}
    \hline\hline
parameters & preliminary values & refitted values\\
\hline
$f_{+,D}^{\pi}(0) \times V_{cd}$ & $0.1420(24)(10)$ & $0.1420(24)(14)$ \\
$r_1$ & $-1.84(22)(7)$ & $-1.85(22)(7)$ \\
$r_2$ & $-1.4(1.5)(0.5)$ & $-1.3(1.5)(0.5)$ \\
\hline\hline
  \end{tabular}
  \caption {\footnotesize Comparison between to be published and refitted values of the
parameters entering in the $z$-expansion parameterization of the hadronic 
form factor $f_{+,D}^{\pi}(q^2)$ using BESIII measurements of 
$D^{0} \rightarrow \pi^- e^+ \nu_e$.}\end{center}
\label{tab:verif_bes3}
\end{table}

\subsection{Combination of all measurements and comparison with HFAG results~\cite{hfag}}

Measurements
from Belle, CLEOc tagged and untagged analyses, BaBar, and BESIII are fitted
using the $z$-expansion parameterization of $f_{+}^{D\pi}(q^2)$.
Values of fitted parameters are given in tab.~\ref{tab:zfit_comp},
  where they are compared with
values computed by HFAG for the average of CLEOIII, CLEO-c tagged and 
untagged
analyses, and using only the total branching fraction measured by Belle.

\begin{table}[!htbp!]
\begin{center}
   \begin{tabular}{lcc}
     \hline\hline
parameter & Belle/CLEO-c/BaBar & HFAG 2012 \\
\hline
$\chi^2/{\rm ndf}$ & $51.1/55$ &$5.5/7$ \\
$f_{+}^{D\pi}(0)\times |V_{cd}|$ & $0.1424\pm0.0019$ &$0.1472\pm0.0045$ \\
$r_1$ & $-1.94\pm0.19$ &$-2.63\pm0.44$ \\
$r_2$  & $-0.6\pm1.2$ & $4.0 \pm3.0$ \\
\hline\hline
   \end{tabular}
   \caption {\footnotesize Values obtained with a fit to all measurements using the 
$z$-expansion parameterization
of $f_{+}^{D\pi}(q^2)$. The result from HFAG is not including BaBar and BESIII preliminary measurements, 
nor the Belle values versus $q^2$ and is obtained by averaging fitted parameters on individual experimental results.}
\end{center}
\label{tab:zfit_comp}
\end{table}

\end{document}